\documentclass[final,onefignum,onetabnum]{siamltex}
\usepackage[utf8]{inputenc} 
\usepackage[english,activeacute]{babel}
\usepackage{amsmath} 
\usepackage{amsfonts,amssymb,amstext}
\usepackage{amssymb,amsmath,babel,graphicx,epsfig} 
\usepackage{hyperref}

\usepackage{lscape} 
\usepackage{rotating} 
\usepackage{bm} 
\usepackage{multirow} 
\usepackage{verbatim} 
\usepackage{booktabs} 
\usepackage{pgf,tikz} 
\usepackage{multirow} 
\usepackage{float}

\newcommand{\ii}{{\bf i}} 
\newcommand{\R}{{\mathbb R}}
\newcommand{\E}{{\mathbb E}}
\newcommand{\ds}{\displaystyle}
\newtheorem{remark}{Remark} 
\def\d{\hbox{d}}

\title{Vibrato and Automatic Differentiation for High Order Derivatives and Sensitivities of Financial Options} 

\author{Gilles Pag\`es \thanks{Laboratoire de Probabilit\'es et Mod\`eles Al\'eatoires, UMR 7599, UPMC, Case 188, 4 pl. de Jussieu, F-75252 Paris Cedex 5, France, {\tt gilles.pages@upmc.fr.}}
\and
Olivier Pironneau \thanks{Laboratoire Jacques Louis Lions, UMR 7598, Case 187, 4 pl. de Jussieu, F-75252 Paris Cedex 5, France, {\tt olivier.pironneau@upmc.fr.} } 
\and
Guillaume Sall \thanks{Laboratoire de Probabilit\'es et Mod\`eles Al\'eatoires, UMR 7599, UPMC, Case 188, 4 pl. de Jussieu, F-75252 Paris Cedex 5, France, {\tt guillaume.sall@upmc.fr.}}
}
\begin{document} 
\maketitle
\begin{abstract}
	{This paper deals with the computation of second  or higher order greeks of financial securities. It combines two methods, Vibrato and automatic differentiation and compares with other methods. We show that this combined technique is faster than standard finite difference, more stable than automatic differentiation of second order derivatives and more general than Malliavin Calculus. We present a generic framework to compute any greeks and present several applications on different types of financial contracts: European and American options, multidimensional Basket Call and stochastic volatility models such as Heston's model. We give also an algorithm to compute derivatives for the Longstaff-Schwartz  Monte Carlo method for American options. We also extend automatic differentiation for second order derivatives of options with non-twice differentiable payoff.}
\end{abstract}



\pagestyle{myheadings}

\section{Introduction}

Due to BASEL III regulations, banks are requested to evaluate the sensitivities of their portfolios every day (risk assessment). Some of these portfolios are huge and sensitivities are time consuming to compute accurately. Faced with the problem of building a software for this task and distrusting automatic differentiation  for non-differentiable functions, we turned to an idea developed by Mike Giles called Vibrato. 

Vibrato at core is a differentiation of a combination of likelihood ratio method and pathwise evaluation. In Giles \cite{Gil08}, \cite{Gil07}, it is shown that the computing time, stability and precision are enhanced compared with numerical differentiation of the full Monte Carlo path.

In many cases, double sensitivities, i.e. second derivatives with respect to parameters, are needed (e.g. gamma hedging). 

Finite difference approximation of sensitivities  is a very simple method but its precision is hard to control because it relies on the appropriate choice of the increment. Automatic differentiation  of computer programs bypass the difficulty and its computing cost is similar to finite difference, if not cheaper.  But in finance the payoff is never twice differentiable and so generalized derivatives have to be used requiring approximations of Dirac functions of which the precision is also doubtful.

The purpose of this paper is to investigate the feasibility of Vibrato for second and higher derivatives. We will  first compare Vibrato applied twice with the analytic differentiation of Vibrato and show that it is equivalent; as the second is easier we propose the best compromise for second derivatives: Automatic Differentiation of Vibrato.

 In \cite{Cap14}, Capriotti has recently investigated the coupling of different mathematical methods -- namely pathwise and likelihood ratio methods -- with an Automatic differentiation technique for the computation of the second order greeks; here we follow the same idea but with Vibrato and also for the computation of higher order derivatives.

Automatic Differentiation (AD) of computer program as described by Greiwank in \cite{Gre89}, \cite{GW08}, Naumann in \cite{Nau12} and Hascoet in \cite{TapenadeRef13} can be used in direct or reverse mode. In direct mode the computing cost is similar to finite difference but with no roundoff errors on the results: the method is exact because every line of the computer program which implements the financial option is differentiated exactly. The computing cost of a first derivative is similar to running the program twice.

Unfortunately, for many financial products the first or the second sensitivities do not exist at some point, such is the case for the standard Digital option at $x=K$; even the payofff of the a plain vanilla European option is not twice differentiatble at $x=K$, yet the Gamma is well defined due to the regularizing effect of the Brownian motion (or the heat kernel) which gives sense to the expectation of a Dirac as a pointwise value of a probability density; in short the end result is well defined but the intermediate steps of AD are not.

We tested \texttt{ADOL-C}  \cite{GW10} and tried to compute the Hessian matrix for a standard European Call option in the Black-Scholes model but the results were wrong.   So we adapted our AD library based on operator overloading by including approximations of Dirac functions and obtained decent results; this is the second conclusion of the paper: AD for second sensitivities can be made to work; it is simpler than Vibrato+AD (VAD) but it is risky and slightly more computer intensive.

More details on AD can be found in Giles et al. \cite{GG05}, Pironneau \cite{Pir08}, Capriotti \cite{Cap11}, Homescu \cite{Hom11} and the references therein.

An important constraint when designing costly software for risk assessment is to be compatible with the history of the company which contracts the software; most of the time, this rules out the use of partial differential equations (see \cite{AP05}) as most quant companies use Monte Carlo algorithms for pricing their portfolios.

For security derivatives computed by a Monte Carlo method, the computation of their sensitivities with respect to a parameter is most easily approximated by finite difference (also known as the {\em shock method}) thus requiring the reevaluation of the security with an incremented parameter. There are two problems with this method: it is imprecise when generalized to higher order derivatives and expensive for multidimensional problems with multiple parameters. The $n^{th}$ derivative of a security with $p$ parameters requires $(n+1)p$ evaluations; furthermore the choice of the perturbation parameter is tricky.

From a semi-analytical standpoint the most natural way to compute a sensitivity is the pathwise method  described in Glasserman \cite{Gla03} which amounts to compute the derivative of the payoff for each simulation path. Unfortunately, this technique happens to be inefficient for certain types of payoffs including some often used in quantitative finance like Digitals or Barrier options. For instance, as it is not possible to obtain the Delta of a Digital Call that way (the derivative of the expectation of a Digital payoff is not equal to the expectation of the derivative of the Digital payoff, which in fact does not exist as a function), the pathwise method cannot evaluate the Gamma of a Call option in a standard Black-Scholes model. The pathwise derivative estimation is also called {\it infinitesimal perturbation} and there is a extensive literature on this subject; see for example Ho et al. \cite{HC83}, in Suri et al. \cite{SZ88} and in L'Ecuyer \cite{Ecu90}. A general framework for some applications to option pricing is given in Glasserman \cite{Gla91}.

There are also two well known mathematical methods to obtain sensibilities, the so-called $\log$-likelihood ratio method and the Malliavin calculus. However, like the pathwise method, both have their own advantage and drawback. For the former, the method consists in differentiating the probability density of the underlying and clearly, it is not possible to compute greeks if the probability density of the underlying is not known. Yet, the method has a great advantage in that the probability densities are generally smooth functions of their parameters, even when payoff functions are not. This method has been developed primarily in Glynn \cite{Gly87}, Reiman et al. \cite{RW89}, Rubinstein \cite{Rub89} and some financial applications in Broadie et al. \cite{BG96} and Glasserman et al. \cite{GZ99}. 

As for the Malliavin calculus, the computation of the greeks consists in writing the expectation of the orignal payoff function times a specific factor i.e. the Malliavin weight which is a Skorohod integral, the adjoint operator of the Malliavin derivative. The main problem of this method is that the computation of the Malliavin weight can be complex and/or computationally costly for a high dimensional problem. Several articles deal with the computation of greeks via Malliavin calculus, Fourni\'e et al. \cite{FLLL01}, Benhamou \cite{Ben03} and Gobet et al. \cite{GE05} to cite a few. The precision of the Malliavin formulae also degenerates for short maturities, especially for the $\Delta$-hedge.

Both the likelihood ratio and the Malliavin calculus are generally faster than the pathwise or finite difference method because, once the terms in front of the payoff function (the weight is computed analytically), the approximation of a greek in a one-dimensional case is almost equivalent to the cost of the evaluation of the pricing function. One systematic drawback is the implementation of these method in the financial industry is limited by the specific analysis required by each new payoff.

\bigskip

The paper is organized as follows; in  section \ref{VIBRATO} we begin by recalling the Vibrato method for first order derivatives as in Giles \cite{Gil08} for the univariate and the multivariate case. We then generalize the method for the second and higher order derivatives with respect to one or several parameters and we describe the coupling to an analytical or Automatic differentiation method to obtain an additional order of differentiation.

In section \ref{AUTODIFF}, we recall briefly the different methods of Automatic differentiation.  We describe the direct and the adjoint or reverse mode to differentiate a computer program. We also explain how it can be extended to some non differentiable functions.

Section \ref{APPLICATION} deals with several applications to different derivative securities. We show some results of second order derivatives (Gamma and Vanna) and third order derivatives in the case of a standard European Call option: the sensitivity of the Gamma with respect to changes in the underlying asset and a cross-derivatives with respect to the underlying asset, the volatility and the interest rate. Also, we compare different technique of Automatic differentiation and we give some details about our computer implementations.

In section \ref{AMERICAN} we study some path-dependent products; we apply the combined Vibrato plus Automatic differentiation method to the computation of the Gamma for an American Put option computed with the Longstaff Schwartz algorithm \cite {LS01}. We also illustrate the method on a multidimensional Basket option (section \ref{APPLICATION}) and on a European Call with Heston's model in  section \ref{HESTON}. 
In section \ref{REVERSEAD}, we study the computing time for the evaluation of the Hessian matrix of a standard European Call Option in the Black-Scholes model. Finally, in  section \ref{MALLIAVIN} we compare VADs to Malliavin's and to the likelihood ratio method in the context of short maturities.

\section{Vibrato}\label{VIBRATO}

Vibrato was introduced by Giles in \cite{Gil08}; it is based on a reformulation of the payoff which is better suited to differentiation. The Monte Carlo path is split into the last time step and its past. Let us explain the method on a plain vanilla multi-dimensional option.

First, let us recall the likelihood ratio method for derivatives. \\
Let the parameter set $\Theta$ be a subset of ${\R}^p$. Let $b:\Theta \times {\R}^d\rightarrow {\R}^d$, $\sigma :\Theta \times {\R}^d \rightarrow {\R}^{d\times q}$ be continuous functions, locally Lipstchitz in the space variable, with linear growth, both uniformly in $\theta \!  \in \Theta$. We omit time as variable in both $b$ and $\sigma$ only for simplicity. And let $(W_t)_{t\geq0}$ be a $q$-dimensional standard Brownian motion defined on a probability space $(\Omega, \mathcal{A}, \mathbb{P})$.

\begin{lemma}
	\label{LRM}{\rm(Log-likelihood ratio)}\\
	
	Let $p(\theta,\cdot)$ be the probability density of a random variable $X(\theta)$, which is function of $\theta$; consider 
	\begin{equation}
		\mathbb{E}[V(X(\theta))]= \int_{{\R}^d}V(y)p(\theta,y)dy.
	\end{equation}
	If $\theta \mapsto p(\theta,\cdot)$ is differentiable at $\theta^0\!\in\Theta$ for all $y$, then, under a standard domination or a uniform integrability assumption one can interchange differentiation and integration : for $i=1,..,p$,

	\begin{equation}
		\frac{
		\partial}{
		\partial \theta_i}\Big[ \mathbb{E}[V(X(\theta))]\Big]_{|\theta^0}=\int_{{\R}^d}V(y) \frac{
		\partial \log{p}}{
		\partial \theta_i}(\theta^0,y)p(\theta^0,y)dy =\mathbb{E}\left[ V(X(\theta)) \frac{
		\partial \log{p}}{
		\partial \theta_i}(\theta,X(\theta))\right]_{|\theta^0}.
	\end{equation}
\end{lemma}

\subsection{Vibrato for a European Contract}\label{VibEuro}

Let $X=(X_t)_{t\in[0,T]}$ be a diffusion process, the strong solution of the following Stochastic Differential Equation (SDE) 
\begin{equation}
	\label{1e} dX_t=b\left(\theta, X_t\right)dt +\sigma(\theta, X_t)dW_t,~~X_0=x. 
\end{equation}
For simplicity and without loss of generality, we assume that $q=d$; so $\sigma$ is a square matrix. Obviously, $X_t$ depends on $\theta$; for clarity, we write $X_t(\theta)$ when the context requires it. \\\\
Given an integer $n>0$, the Euler scheme with constant step $h=\frac{T}{n}$, defined below in~(\ref{1e}), approximates $X_t$ at time $t_{k}^n=k h$ , i.e. $\bar{X}_{k}^n\approx X_{kh}$,  and it is recursively defined by 
\begin{equation}
	\label{1eb} \bar{X}_{{k}}^n=\bar{X}_{{k-1}}^n+b(\theta,\bar{X}_{{k-1}}^n)h+\sigma(\theta, \bar{X}_{k-1}^n)\sqrt{h}Z_{k},~~\bar X^n_0=x,~~k=1,\dots,n, 
\end{equation}
where $\{{Z_k}\}_{k=1,..,n}$ are independent random Gaussian ${\cal N}(0,I_{d})$ vectors. The relation between $W$ and $Z$ is 
\begin{equation}
	W_{{t^n_k}}-W_{t^n_{k-1}}=\sqrt{h}Z_k. 
\end{equation}

Note that $\bar{X}_{n}^n=\mu_{n-1} (\theta)+\sigma_{n-1}(\theta) \sqrt{ h } Z_n$ with 
\begin{equation}
	\label{sigmu} \mu_{n-1} (\theta)=\bar{X}^n_{{n-1}}(\theta)+b(\theta, \bar{X}^n_{{n-1}}(\theta) )h\;\mbox{ and }\;\sigma_{n-1}(\theta)=\sigma ( \theta, \bar{X}^n_{{n-1}}(\theta))\sqrt{h}. 
\end{equation}
Then, for any Borel function $V:\R^d\to\R$ such that $\mathbb{E}|V(\bar{X}^n_{n}(\theta))|<+\infty$, 
\begin{equation}
	\label{2e} \mathbb{E}\left[V(\bar{X}^n_{n}(\theta))\right]=\mathbb{E}\left[\mathbb{E}\left[V(\bar{X}^n_{n}(\theta)) \mid (W_{t^n_{k}})_{k=0,\dots,n-1}\right]\right] =\mathbb{E}\left[\mathbb{E}\left[V(\bar{X}^n_{n}(\theta))\mid \bar{X}^n_{{n-1}}\right]\right]. 
\end{equation}
This follows from the obvious fact that the Euler scheme defines a Markov chain $\bar X$ with respect to the filtration ${\cal F}_k= \sigma(W_{t^n_\ell}, \, \ell=0,\dots,k)$.

Furthermore, by homogeneity of the chain, 
\begin{equation}
	\label{1e2} \mathbb{E}\left[V(\bar{X}^n_{n}(\theta))\mid \bar{X}^n_{{n-1}}\right]=\left\{ \mathbb{E}_x \left[V(\bar X^n _1(x,\theta))\right] \right\}_{\left| { x=\bar{X}^n_{{n-1}}}\right.} =\left\{\E[V(\mu + \sigma \sqrt h Z) ] \right\} _{ \tiny 
	\left|\begin{matrix}
		\mu = \mu_{n-1}(\theta) \\
		\sigma = \sigma _{n-1}(\theta) 
	\end{matrix}\right.
	} .
\end{equation}
Where $\bar X^n _1(x,\theta)$ denotes the value at time $t^n_1$ of the Euler scheme with $k=1$, starting at $x$  and where the last expectation is with respect to $Z$.

\subsection{First Order Vibrato}

We denote $\varphi(\mu, \sigma )=\mathbb{E}\left[ V(\mu+\sigma \sqrt h Z) \right]$. From (\ref{2e}) and (\ref{1e2}), for any $i\in(1,\dots,p)$ 
\begin{equation}
	\label{vb1c} \frac{
	\partial}{
	\partial \theta_i}\mathbb{E}[V( \bar X_n^n(\theta) ) ] =\displaystyle \mathbb{E}\left[ \frac{
	\partial}{
	\partial \theta_i } \left\{\mathbb{E}[V(\mu + \sigma \sqrt h Z) ] \right\} _{\tiny 
	\left|\begin{matrix}
		\mu = \mu_{n-1}(\theta) \\
		\sigma = \sigma _{n-1}(\theta) 
	\end{matrix}\right.
	} \right] = \displaystyle \mathbb{E}\left[ \frac{
	\partial\varphi }{
	\partial \theta_i }( \mu_{n-1}(\theta), \sigma _{n-1}(\theta) ) \right]
\end{equation}
and 
\begin{eqnarray}
	 \label{vib1dd} \ds\frac{
	\partial \varphi }{
	\partial \theta_i } \left( {\mu_{n-1}}, \sigma_{n-1} \right) 
    \displaystyle =\frac{
	\partial {\mu_{n-1}}}{
	\partial \theta_i}\cdot \frac{
	\partial \varphi}{
	\partial {\mu}}({\mu_{n-1}},\sigma_{n-1}) +\frac{
	\partial \sigma_{n-1}}{
	\partial \theta_i} :\frac{
	\partial \varphi}{
	\partial \sigma}({\mu_{n-1}},\sigma_{n-1}) 
\end{eqnarray}
where $\cdot$ denotes the scalar product and $:$ denotes the trace of the product of the matrices.
\begin{lemma}
	
	The ${\theta_i}$-tangent process to $X$,
	$\ds Y_t=\frac{
			\partial X_t}{
			\partial {\theta_i}}$,
	  is defined as the solution of the following $SDE$ (see Kunita\cite{Kun90} for a proof) 
	\begin{equation}
		\label{3e} 
\d Y_t=\left[ b'_{\theta_i}(\theta,X_t)+ b'_x(\theta,X_t) Y_t \right]\d t+\left[ \sigma'_ {\theta_i}(\theta,X_t) + \sigma'_x (\theta,X_t) Y_t \right]\d W_t,~~
Y_0=\frac{\partial X_0}{\partial {\theta_i}}
	\end{equation}
	where the primes denote standard derivatives.
\end{lemma}
As for  $\bar X^n_k$ in (\ref{1e}), we may discretize (\ref{3e}) by 
\begin{eqnarray}
	\label{3ee} \bar Y^n_{{k+1}}&=&\bar Y^n_{{k}}+ \left[ b '_{\theta_i} (\theta, \bar X^n_{{k}}) +b '_x(\theta, \bar X^n_{{k}})\bar Y^n_{{k}}\right]h 	
	+ \left[ \sigma '_{\theta_i} (\theta, \bar X^n_{{k}}) + \sigma '_x(\theta, \bar X^n_{{k}}) \bar Y^n_{{k}}\right]\sqrt h Z_{k+1}.
\end{eqnarray}
Then from (\ref{sigmu}), 
\begin{eqnarray}\label{defMUSIG}
	\label{vbmu}&& \frac{
	\partial \mu _{n-1}}{
	\partial {\theta_i}} = \bar Y^n_{{n-1}}(\theta) + h\left[ b'_{\theta_i}(\theta,\bar X ^n_{{n-1}}(\theta))+ b'_x(\theta,\bar X ^n_{{n-1}}(\theta))\bar Y^n_{{n-1}}(\theta)\right] \cr&& \frac{
	\partial \sigma _{n-1}}{
	\partial {\theta_i}} = \sqrt{h}\left[ \sigma'_{\theta_i}(\theta,\bar X ^n_{{n-1}}(\theta))+ \sigma'_x(\theta,\bar X ^n_{{n-1}}(\theta))\bar Y^n_{{n-1}}(\theta)\right]. 
\end{eqnarray}
So far we have shown the following lemma.
\begin{lemma}\label{lemma3}
	When $X_n^n(\theta)$ is given by (\ref{1e}), then $\displaystyle\frac{
	\partial}{
	\partial \theta_i}\mathbb{E}[V( \bar X_n^n(\theta) ) ] $ is given by (\ref{vb1c}) with (\ref{vib1dd}), (\ref{vbmu}) and (\ref{3ee}). 
\end{lemma}

In (\ref{1e}) $b$ and $\sigma$ are constant in the time interval $(k h, (k+1)h)$, therefore the conditional probability of $\bar{X}^n_{n}$ given $\bar{X}^n_{{n-1}}$ given by
\begin{equation}
	\label{d1d} p(x)= \frac{1}{(\sqrt{2\pi})^d\sqrt{\left| {\Sigma}\right|}}e^{-\frac{1}{2}(x-{\mu})^T{\Sigma}^{-1}(x-{\mu})} 
\end{equation}
where $\mu$ and $\Sigma=h\sigma\sigma^T$ are evaluated at time $(n-1)h$ and given by (\ref{sigmu}). As in Dwyer et al. \cite{DM48}, 
\begin{eqnarray*}
	& \displaystyle \frac{
	\partial}{
	\partial\mu}\log p(x) ={\Sigma}^{-1}(x-{\mu}),~~ &
	\displaystyle \frac{
	\partial}{
	\partial\Sigma}\log p(x)=-\frac{1}{2}\Sigma ^{-1}+\frac{1}{2} \Sigma ^{-1} (x-{\mu})(x-{\mu})^T \Sigma ^{-1} 
	\Rightarrow ~
	\cr&
	\displaystyle \frac{
	\partial}{
	\partial\mu}\log p(x)|_{x=X_n^n} = {\sigma}^{-T}\frac{Z}{\sqrt{h}},  & ~ ~\displaystyle \frac{
	\partial}{
	\partial\Sigma}\log p(x)|_{x=X_n^n} =\frac{1}{2h} \sigma ^{-T}( Z Z ^T - I ) \sigma^{-1}. 
\end{eqnarray*}
Finally, applying Lemma \ref{lemma3} and Lemma \ref{LRM} yields the following proposition

\begin{theorem}
	{{\rm (Vibrato, multidimensional first order case)}}
	
	\begin{equation}\label{vib1}
		\begin{aligned}
			\frac{
			\partial }{
			\partial \theta_i }\mathbb{E}[V( \bar X_n^n(\theta) ) ] &=
			\displaystyle \mathbb{E}\left[ \frac{
	\partial}{
	\partial \theta_i } \left\{\mathbb{E}[V(\mu + \sigma \sqrt h Z) ] \right\} _{\tiny 
	\left|\begin{matrix}
		\mu = \mu_{n-1}(\theta) \\
		\sigma = \sigma _{n-1}(\theta) 
	\end{matrix}\right.
	} \right] 
	\cr&
	=\mathbb{E}\left[\frac{1}{\sqrt{h}} \frac{
			\partial \mu}{
			\partial \theta_i} \cdot \mathbb{E}\left[ V(\mu+\sigma \sqrt h Z)\sigma ^{-T}Z \right] \right. \left|_{\tiny 
			\left|\begin{matrix}
				\mu = \mu_{n-1}(\theta) \\
				\sigma = \sigma _{n-1}(\theta) 
			\end{matrix}\right.
			} \right.
			\cr & 
			+\frac{1}{2h}\left. \left. \frac{
			\partial \Sigma}{
			\partial \theta_i}:\mathbb{E}\left[ V(\mu+\sigma \sqrt h Z) \sigma ^{-T}( Z Z ^T - I ) \sigma ^{-1} \right]  \right|_{\tiny 
			\left|\begin{matrix}
				\mu = \mu_{n-1}(\theta) \\
				\sigma = \sigma _{n-1}(\theta) 
			\end{matrix}\right.
			} \right]
		\end{aligned}.
	\end{equation}
\end{theorem}

\subsection{Antithetic Vibrato}
One can expect to improve the above formula -- that is, reducing its variance -- by the means of antithetic transform (see section \ref{RVVAD} below for a short discussion)
	\label{VUC} The following holds:
		\begin{equation}
		\mathbb{E}\left[ V(\mu+\sigma \sqrt h Z)\sigma ^{-T}Z \right] = \frac{1}{2}\mathbb{E}\left[\left( V(\mu+\sigma \sqrt h Z) - V(\mu-\sigma \sqrt h Z) \right) \sigma^{-T}Z \right].
	\end{equation}
	similarly, using $E[ZZ^T-I] = 0$, 
	\begin{eqnarray}
		&& \mathbb{E}\left[ V(\mu+\sigma \sqrt h Z) \sigma ^{-T}( Z Z ^T - I ) \sigma ^{-1} \right] \cr&& = \frac{1}{2}\mathbb{E}\left[ \left( V(\mu+\sigma \sqrt h Z) -2V(\mu) +V(\mu-\sigma \sqrt h Z)\right)\sigma ^{-T}( Z Z ^T - I) \sigma ^{-1}\right].
	\end{eqnarray}

\begin{corollary}
	{\rm (One dimensional case, d=1)}
	
	\begin{eqnarray}
		\label{d1}&\displaystyle \frac{
		\partial }{
		\partial \theta_i }&\mathbb{E}[V( \bar X_n^n(\theta) ) ] = \frac12\mathbb{E}\left[ \frac{
		\partial \mu}{
		\partial \theta_i} \mathbb{E}\left[ \left( V(\mu+\sigma\sqrt h Z) - V(\mu-\sigma\sqrt h Z) \right) \frac{Z}{\sigma\sqrt h } \right] \right.\left|_{\tiny 
		\left|\begin{matrix}
			\mu = \mu_{n-1}(\theta) \\
			\sigma = \sigma _{n-1}(\theta) 
		\end{matrix}\right.
		}\right. 
\cr&& 
		+ \left.\left.\frac{
		\partial \sigma}{
		\partial \theta_i}\mathbb{E}\left[ \left( V(\mu+\sigma\sqrt h Z) -2V(\mu) +V(\mu-\sigma\sqrt h Z)\right)\frac{Z^2-1}{\sigma\sqrt h  } \right]  \right|_{\tiny 
		\left|\begin{matrix}
			\mu = \mu_{n-1}(\theta) \\
			\sigma = \sigma _{n-1}(\theta) 
		\end{matrix}\right.
		} \right]
	\end{eqnarray}
\end{corollary}
\paragraph{Conceptual Algorithm} 
In figure \ref{vibscheme} we have illustrated the Vibrato decomposition at the path level.  To implement the above one must perform the following steps:
\begin{figure}
	[htbp] 
	\begin{center}
		\includegraphics[width=0.49
		\textwidth]{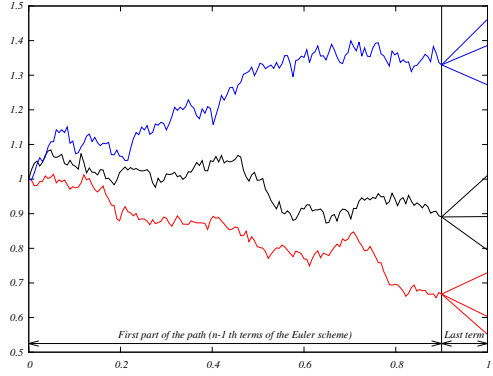} \caption{\label{vibscheme}{\it Scheme of simulation path of the Vibrato decomposition.} }
	\end{center}
\end{figure}
\begin{enumerate}
	\item Choose the number of time step $n$, the number of Monte-Carlo path $M$ for the $n-1$ first time steps, the number $M_Z$ of replication variable $Z$ for the last time step. 
	\item For each Monte-Carlo path $j=1..M$ 
	\begin{itemize}
		\item Compute $\{X^n_k\}_{k=1:n-1}$, $\mu_{n-1},\sigma_{n-1}$ by (\ref{1e}),			(\ref{sigmu}). 
	\item Compute $V(\mu_{n-1})$
	\item Compute $\displaystyle\frac{\partial\mu_{n-1}}{\partial\theta_i}$ and $ \displaystyle\frac{\partial\sigma_{n-1}}{\partial\theta_i}$ by  (\ref{3e}), (\ref{vbmu}) and (\ref{3ee})		\item Replicate $M_Z$ times the  last time step, i.e. 
		
		For $m_Z\in(1,\ldots,M_Z)$ 
		\begin{itemize}
			\item Compute $V(\mu_{n-1}+\sigma_{n-1}\sqrt h Z^{(m_Z)})$ and $V(\mu_{n-1}-\sigma_{n-1}\sqrt h Z^{(m_Z)})$  
			\end{itemize}
		\end{itemize}	
		\item	
		  In (\ref{d1}) compute the inner expected value by averaging over all $M_Z$ results, then multiply by $\frac{\partial\mu}{\partial\theta_i}$ and $\frac{\partial\sigma}{\partial\theta_i}$ and then average over the $M$ paths.
\end{enumerate}
\begin{remark}
For simple cases such as of the sensibilities of European options, a small $M_Z$ suffices; this is because there is another average with respect to M in the outer loop.
\end{remark}
\begin{remark}
For European options one may also use the Black-Scholes formula for the expected value in (\ref{vib1}).
\end{remark}

\subsection{Second Derivatives}
Assume that $X_0$, $b$ and $\sigma$ depend on two parameters $(\theta _1, \theta_2 ) \in \Theta^2$.		There are two ways to compute second order derivatives. Either by differentiating the Vibrato (\ref{vib1}) while using Lemma \ref{LRM} or by applying the Vibrato idea to the second derivative. 
\subsubsection{Second Derivatives by Differentiation of Vibrato}
Let us differentiate (\ref{vib1}) with respect to a second parameter $\theta_j$:
	\begin{equation}\label{vibb1}
		\begin{aligned}
			\frac{
			\partial^2 }{
			\partial \theta_i\partial\theta_j }\mathbb{E}[V(X_T ) ] &= \mathbb{E}\left[\frac{1}{\sqrt{h}}\Big( \frac{
			\partial^2 \mu}{
			\partial \theta_i\partial \theta_j} \cdot \mathbb{E}\left[ V(\mu+\sigma \sqrt h Z)\sigma ^{-T}Z \right] \right.
\cr&  \left.
			+  \frac{
			\partial \mu}{
			\partial \theta_i} \cdot \frac{\partial}{
			\partial \theta_j}\mathbb{E}\left[ V(\mu+\sigma \sqrt h Z)\sigma ^{-T}Z \right] 
			\Big)\right. \left|_{\tiny 
			\begin{matrix}
				\mu = \mu_{n-1}(\theta) \\
				\sigma = \sigma _{n-1}(\theta) 
			\end{matrix}
			} \right.
			\cr & 
			+\frac{1}{2h}\Big( \frac{
			\partial^2 \Sigma}{
			\partial \theta_i\partial\theta_j}:\mathbb{E}\left[ V(\mu+\sigma \sqrt h Z) \sigma ^{-T}( Z Z ^T - I ) \sigma ^{-1} \right] 
		\cr & 
			+\left.\left. \frac{
			\partial \Sigma}{
			\partial \theta_i}:\frac{\partial}{
			\partial \theta_j}\mathbb{E}\left[ V(\mu+\sigma \sqrt h Z) \sigma ^{-T}( Z Z ^T - I ) \sigma ^{-1} \right] \Big) \right|_{\tiny 
			\begin{matrix}
				\mu = \mu_{n-1}(\theta) \\
				\sigma = \sigma _{n-1}(\theta) 
			\end{matrix}
			} \right]
		\end{aligned}
	\end{equation}
	The derivatives can be expanded further; for instance in the one dimensional case and after a tedious algebra one obtains:
\begin{theorem}{\rm (Second Order by Differentiation of Vibrato)} \label{propVAD}
{\small			\begin{eqnarray}
			\label{vibvibad} &&\ds
						\frac{\partial^2 }{\partial \theta^2 } \mathbb{E}[V(X_T)]
			=
			 \mathbb{E}\left[ \frac{
			\partial ^2 \mu }{
			\partial \theta ^2}\mathbb{E}\left[ V(\mu+\sigma \sqrt h Z)\frac{Z}{\sigma \sqrt{h}} \right] +\left(\frac{
			\partial \mu}{
			\partial \theta}\right)^2\mathbb{E}\left[ V(\mu+\sigma \sqrt h Z)\frac{Z^2-1}{\sigma^2 h} \right] 
			\right.\cr&& \left. 
			+\left(\frac{
			\partial \sigma}{
			\partial \theta}\right)^2\mathbb{E}\left[ V(\mu+\sigma \sqrt h Z)
			\frac{Z^4-5Z^2+2}{\sigma ^2 h} \right] 
			\right.	 \cr&& \left. 
			+ \frac{
			\partial ^2\sigma }{
			\partial \theta ^2}\mathbb{E}\left[ V(\mu+\sigma \sqrt h Z)\frac{Z^2-1}{\sigma \sqrt{h}} \right] +2 \frac{
			\partial \mu }{
			\partial \theta }\frac{
			\partial \sigma}{
			\partial \theta }\mathbb{E}\left[V(\mu+\sigma \sqrt h Z)\frac{Z^3-3Z}{\sigma^2 h} \right] \right] 
		\end{eqnarray}
}
\end{theorem}
\subsubsection{Second Derivatives by Second Order Vibrato}
The same Vibrato strategy can be applied also directly to second derivatives.
		
		As before the derivatives are transfered to the PDF $p$ of $X_T$:
\begin{eqnarray}&
\ds \frac{\partial^2}{\partial\theta_i\partial\theta_j}\E[V(X_T)]
&= \int_{\R^d}\frac{V(x)}{p(x)}\frac{\partial^2 p}{\partial\theta_i\partial\theta_j}p(x)\d x
=  \int_{\R^d}V(x)[\frac{\partial^2\ln p}{\partial\theta_i\partial\theta_j}+ \frac{\partial\ln p}{\partial\theta_i}\frac{\partial\ln p}{\partial\theta_j} ]p(x)\d x
\cr&&
= \E\left[V(x)\left(\frac{\partial^2\ln p}{\partial\theta_i\partial\theta_j}+ \frac{\partial\ln p}{\partial\theta_i}\frac{\partial\ln p}{\partial\theta_j} \right)\right]
\end{eqnarray}
Then 
		\begin{eqnarray*}
			\frac{
			\partial ^2}{
			\partial \theta _1 
			\partial \theta _2 }\mathbb{E}[V(\bar X ^n_T(\theta _1, \theta _2))] &=& \frac{
			\partial ^2 \varphi}{
			\partial \theta _1 
			\partial \theta _2 }(\mu, \sigma ) \cr &=& \frac{
			\partial \mu }{
			\partial \theta _1} \frac{
			\partial \mu }{
			\partial \theta _2} \frac{
			\partial ^2 \varphi}{
			\partial \mu ^2}(\mu , \sigma ) + \frac{
			\partial \sigma }{
			\partial \theta _1} \frac{
			\partial \sigma }{
			\partial \theta _2}\frac{
			\partial ^2 \varphi}{
			\partial \sigma^2}(\mu , \sigma ) + \frac{
			\partial ^2 \mu }{
			\partial \theta_1
			\partial \theta_2} \frac{
			\partial \varphi}{
			\partial \mu}(\mu , \sigma ) \cr && + \frac{
			\partial ^2 \sigma }{
			\partial \theta_1
			\partial \theta_2} \frac{
			\partial \varphi}{
			\partial \sigma}(\mu , \sigma ) +\left( \frac{
			\partial \mu }{
			\partial \theta_1} \frac{
			\partial \sigma }{
			\partial \theta_2} + \frac{
			\partial \sigma }{
			\partial \theta _1} \frac{
			\partial \mu }{
			\partial \theta_2} \right)\frac{
			\partial ^2 \varphi }{
			\partial \mu 
			\partial \sigma}(\mu , \sigma). 
		\end{eqnarray*}
		We need to calculate the two new terms $\displaystyle \frac{
		\partial ^2}{
		\partial \theta _1 
		\partial \theta_2}\mu_{n-1}(\theta _1, \theta _2)$ and $\displaystyle \frac{
		\partial ^2}{
		\partial \theta _1 
		\partial \theta_2}\sigma _{n-1}(\theta _1, \theta _2)$. It requires the computation of the first derivative with respect to $\theta_i$ of the tangent process $Y_t$, that we denote $Y^{(2)}_t(\theta_1, \theta_2)$. 
		
		Then (\ref{vbmu}) is differentiated and an elementary though tedious computations yields the following proposition:
		%
		\begin{proposition}
			\label{P3} 
			
			The $\theta_i$-tangent process $Y^{(i)}_{t}$ defined above in Lemma \ref{3e} has a $\theta_j$-tangent process $Y^{(ij)}_t$ defined by 
			\begin{eqnarray}
				\nonumber dY^{(ij)}_t &=& \left[ 
				b''_{ \theta _i\theta_j} (\theta _1,\theta _2,X_t)  
				+  b''_{ \theta _i, x} (\theta _1,\theta _2,X_t)Y^{(j)}_{t}
				+   b''_{ \theta _j, x} (\theta _1,\theta _2,X_t)Y^{(i)}_{ t} 
				\right. \\
				\nonumber && \left.  + b''_{x^2}(\theta _1,\theta _2,X_t)Y^{(i)}_{ t}Y^{(j)}_{ t}
				+b'_x(\theta _1, \theta _2,X_t)Y^{(ij)}_t
				\right] \d t \\
				\nonumber &&+ \left[ \sigma''_{ \theta _i\theta_j} (\theta _1,\theta _2,X_t)  
				+  \sigma''_{ \theta _i, x} (\theta _1,\theta _2,X_t)Y^{(j)}_{t}
				+   \sigma''_{ \theta _j, x} (\theta _1,\theta _2,X_t)Y^{(i)}_{ t} 
				\right. \\
				\nonumber && \left.  + \sigma''_{x^2}(\theta _1,\theta _2,X_t)Y^{(i)}_{ t}Y^{(j)}_{ t}
				+\sigma'_x(\theta _1, \theta _2,X_t)Y^{(ij)}_t
				 \right]dW_t. 
			\end{eqnarray}
		\end{proposition}

		Finally in the univariate case $\theta=\theta_1=\theta_2$ this gives 

\begin{proposition}{\rm (Second Order Vibrato)}\label{propVIBVIB}
{\small			\begin{eqnarray}
			\label{vibadga3} &&\ds
						\frac{\partial^2 }{\partial \theta^2 } \mathbb{E}[V(X_T)]
			=
			\cr&&
			 \mathbb{E}\left[ \frac{
			\partial ^2 \mu }{
			\partial \theta ^2}\mathbb{E}\left[ V(\mu+\sigma \sqrt h Z)\frac{Z}{\sigma \sqrt h } \right] +\left(\frac{
			\partial \mu}{
			\partial \theta}\right)^2\mathbb{E}\left[ V(\mu+\sigma \sqrt h Z)\frac{Z^2-1}{\sigma^2 h} \right] +\left(\frac{
			\partial \sigma}{
			\partial \theta}\right)^2\mathbb{E}\left[ V(\mu+\sigma \sqrt h Z)\right.	\right. \cr&& \left. \left.
			\frac{Z^4-5Z^2+2}{\sigma ^2 h} \right] + \frac{
			\partial ^2\sigma }{
			\partial \theta ^2}\mathbb{E}\left[ V(\mu+\sigma \sqrt h Z)\frac{Z^2-1}{\sigma \sqrt{h}} \right] +2 \frac{
			\partial \mu }{
			\partial \theta }\frac{
			\partial \sigma}{
			\partial \theta }\mathbb{E}\left[V(\mu+\sigma \sqrt h Z)\frac{Z^3-3Z}{\sigma^2 h} \right] \right] 
		\end{eqnarray}
}
%
\end{proposition}		

\begin{remark}
It is equivalent to Proposition \ref{propVAD} hence to the direct differentiation of Vibrato.
\end{remark}	

		\subsection{Higher Order Vibrato} The Vibrato-AD method can be generalized  to higher order of differentiation of Vibrato with respect to the parameter $\theta$ with the help of the Fa\`a di Bruno formula and its generalization to a composite function with a vector argument, as given in Mishkov \cite{Mis99}.
		
		\subsection{Antithetic Transform, Regularity and Variance}\label{RVVAD}
In this section, we assume $d=q=1$ for simplicity.

Starting from Vibrato $\varphi(\mu,\sigma)=\mathbb E [ f (\mu+\sigma  \sqrt{h}Z) ]$ and assuming $f$  Lipschitz continuous with Lipschitz coefficients $[f]_{\text{Lip}}$, we have
{\small
\begin{equation} \label{reduc1}
\displaystyle \frac{\partial \varphi}{\partial \mu}(\mu,\sigma) = \mathbb{E}\left[f(\mu+\sigma  \sqrt{h}Z)\frac{Z}{ \sigma \sqrt{h}}\right] = \mathbb{E}\left[\left(f(\mu+\sigma \sqrt{h}Z) - f(\mu-\sigma Z \sqrt{h})\right)\frac{Z}{2 \sigma \sqrt{h}}\right].
\end{equation}
}
Therefore the variance satisfies
{\small
\begin{eqnarray} \nonumber
\mathbf{Var}\left[\left(f(\mu+\sigma  \sqrt{h}Z)-f(\mu-\sigma  \sqrt{h}Z)\right)\frac{Z}{2\sigma\sqrt{h}}\right]
 &\leq& 
 \mathbb{E} \left[ \left| \left(f(\mu+\sigma  \sqrt{h}Z)-f(\mu-\sigma  \sqrt{h}Z)\right)\frac{Z}{2\sigma\sqrt{h}}\right|^2 \right] 
\cr &
\leq&
 [f]^2_{\text{Lip}}\mathbb{E} \left[ \frac{(2\sigma  \sqrt{h}Z)^2}{4\sigma ^2h} Z^2 \right]
= [f]^2_{\text{Lip}}\mathbb{E}[Z^4]=3[f]^2_{\text{Lip}}.
\end{eqnarray}
}
As $\mathbb E[Z]= 0$, we also have
{\small
\begin{equation} \label{reduc2}
\frac{\partial \varphi}{\partial \mu}(\mu,\sigma) = \mathbb{E} \left[\left( f(\mu+\sigma  \sqrt{h}Z) - f(\mu) \right)\frac Z {\sigma\sqrt{h}} \right].
\end{equation}
}
Then,
{\small
\begin{eqnarray} \nonumber
\mathbf{Var}\left[\left(f(\mu+\sigma  \sqrt{h}Z)-f(\mu)\right)\frac{Z}{ \sigma\sqrt{h}}\right] &\leq& \mathbb{E} \left[ \left| \left(f(\mu+\sigma  \sqrt{h}Z)-f(\mu)\right)\frac{Z}{ \sigma\sqrt{h}}\right|^2 \right] 
\cr &
\leq& 
\frac{1}{\sigma^2h}[f]^2_{\text{Lip}}\mathbb{E} \left[ ( \sigma \sqrt{h}Z)^2 Z^2 \right] = [f]^2_{\text{Lip}}\mathbb{E}[Z^4]=3[f]^2_{\text{Lip}}
\end{eqnarray}
}
\begin{remark}
The variances of formulae (\ref{reduc1}) and (\ref{reduc2}) are equivalent but the latter  is  less expensive to compute. 
\end{remark}
 If $f$ is differentiable and $f'$ has polynomial growth, we also have
\begin{equation}\label{reduc3}
\frac{\partial \varphi}{\partial \mu}(\mu,\sigma)= \mathbb{E}[f'(\mu+\sigma  \sqrt{h}Z)].
\end{equation}
Thus,
{\small
\begin{eqnarray} \nonumber
\mathbf{Var}\left[ f'(\mu+\sigma  \sqrt{h}Z) \right] \leq  \mathbb E \left[ \left( f'(\mu+\sigma \sqrt h Z) \right)^2 \right] \leq& \| f'\|^2_\infty.
\end{eqnarray}
}
\begin{remark} 
Let $f]_\text{Lip}$ denote the Lipschitz constant of $f$.
If $f'$ is bounded, we have $[f]_\text{Lip} = \| f' \| _\infty$ then the expression in (\ref{reduc3}) has a smaller variance than (\ref{reduc1}) and (\ref{reduc2}). 
\end{remark}
Assume that $f'$ is Lipschitz continuous with Lipschitz coefficients $[f']_\text{Lip}$. We can improve the efficiency of (\ref{reduc3}) because
\begin{eqnarray} \nonumber
\mathbf{Var}\left[ f'(\mu+\sigma  \sqrt h Z) \right] &=& \mathbf{Var}\left[ f'(\mu+\sigma  \sqrt h Z ) - f'(\mu) \right] 
\cr & \leq& 
\mathbb{E}\left[ \left| f'(\mu+\sigma  \sqrt h Z) - f'(\mu)\right|^2 \right]
 \leq  [f']^2_\text{Lip}h\sigma^2 \mathbb{E}[Z^2]
 \leq  [f']_\text{Lip} h\sigma ^2
\end{eqnarray}

\begin{remark}
Assuming that $f(x) = \mathbf{1}_{\{x\leq K\} }$, clearly we cannot differentiate inside the expectation and the estimation of the variance seen previously can not be applied.
\end{remark}
\subsubsection{Indicator Function}
Let us assume that $f(x) = \mathbf{1}_{\{x\leq K\} }$.
To simplify  assume that $K\leq\mu$, we have
\begin{equation}\nonumber
\left|f(\mu+\sigma  \sqrt h Z ) - f(\mu - \sigma  \sqrt h Z ) \right| =\left|
 \mathbf{1}_{\left\{Z\leq \frac{K-\mu}{\sigma \sqrt h }\right\}} - \mathbf{1}_{\left\{Z\geq \frac{\mu-K}{\sigma \sqrt h } \right\}}\right|
= \mathbf{1}_{\left\{Z\notin \left[ \frac{K-\mu}{ \sigma \sqrt h },\frac{\mu-K}{\sigma \sqrt h } \right] \right\}},
\end{equation}
hence \begin{equation}\nonumber
\left| \left( f(\mu+\sigma  \sqrt h Z ) - f(\mu - \sigma  \sqrt h Z ) \right) \frac{Z}{ \sigma \sqrt h } \right| =\frac{1}{\sigma \sqrt h } |Z| \mathbf{1}_{\left\{Z\notin \left[ \frac{K-\mu}{\sigma \sqrt h },\frac{\mu-K}{\sigma \sqrt h } \right]\right\}}.
\end{equation}
For the variance, we have
{\small
\begin{eqnarray}&& \nonumber
\mathbf{Var}\left[\left( f(\mu+\sigma \sqrt h Z ) - f( \mu - \sigma \sqrt h Z ) \right)\frac{Z}{ \sigma \sqrt h } \right] 
\cr&&\leq  \mathbb E \left[ \left| \left( f(\mu + \sigma \sqrt h Z ) - f( \mu - \sigma \sqrt h Z ) \right) \frac{Z}{ \sigma \sqrt h } \right|^2\right].
\end{eqnarray}
By Cauchy-Schwarz we can write
\begin{eqnarray}&& \nonumber
\mathbb E \left[ \left| \left( f(\mu + \sigma  \sqrt h Z ) - f( \mu - \sigma  \sqrt h Z ) \right) \frac{Z}{ \sigma \sqrt h} \right|^2\right]
\cr&&=\frac{1}{2\sigma ^2h}\mathbb E\left[ Z^2 \left| f(\mu + \sigma \sqrt h Z ) - f( \mu - \sigma  \sqrt h Z )  \right|^2 \right]
= \frac{1}{2\sigma^2h}\mathbb E \left[ Z^2 \mathbf{1}_{\left\{Z\notin \left[\frac{K-\mu}{ \sigma \sqrt h },\frac{\mu-K}{\sigma \sqrt h }\right]\right\}} \right]
\cr&&
\leq \frac{1}{2\sigma ^2h}\left(\mathbb E [ Z^4] \right)^{\frac{1}2}\left(\mathbb{P}\left(Z\notin\left[\frac{K-\mu}{\sigma \sqrt h },\frac{\mu-K}{ \sigma \sqrt h }\right] \right) \right)^{\frac{1}2}
\leq \frac{\sqrt{3}}{2\sigma ^2h}\left( 2\mathbb P\left( Z \geq \frac{\mu- K}{\sigma \sqrt h }\right)\right)^{\frac12}.
\end{eqnarray}
}
Then
{\small
\begin{eqnarray} \nonumber
\frac{\sqrt{3}}{2\sigma ^2 h}\left( 2\mathbb P\left( Z \geq \frac{\mu- K}{ \sigma \sqrt h }\right)\right)^{\frac12}
= 
\frac{\sqrt 6}{2\sigma ^2 h}\left( \int _{\frac{\mu - K}{\sigma \sqrt h }}^{+\infty} e^{-\frac{u^2}{2}}\frac{du}{\sqrt{2\pi}} \right)^{\frac12}.
\end{eqnarray}
}
Now, $\displaystyle \forall\   a > 0,\   \mathbb P(Z\geq a) \leq \frac{e^{-\frac{a^2}{2}}}{a\sqrt{2\pi}}\ $, so when $a\rightarrow +\infty$,
{\small
\begin{eqnarray}&& \nonumber
\mathbf{Var}\left[ \left( f(\mu+\sigma  \sqrt h Z ) - f( \mu - \sigma  \sqrt h Z ) \right)\frac{Z}{ \sigma \sqrt h } \right]\leq \frac{1}{\sigma^2 h}\sqrt{\frac{3}{2}}\frac{e^{-\frac{(\mu-K)^2}{4\sigma^2 h}}}{(2\pi)^{\frac14}\sqrt{\frac{\mu-K}{ \sigma \sqrt h }}}\cr&& \leq
{{\frac{1}{(2\pi)^{\frac14}\sigma^{\frac32}h^{\frac34}}\sqrt{\frac32}\frac{e^{-\frac{(\mu-K)^2}{4\sigma^2 h}}}{\sqrt{\mu -K}}}}\underset{\sigma \rightarrow 0}{\longrightarrow }
\left \{
\begin{array}{c c}
    0 & \text{if } \mu \neq K \\
    +\infty & \text{otherwise.} \\
\end{array}
\right.
\end{eqnarray}
}
The fact that such estimate can be obtained with non differentiable $f$ demonstrates the power of  the Vibrato technique.
\section{Second Derivatives by Vibrato plus Automatic Differentiation (VAD)} \label{AUTODIFF}
		
			The differentiation that leads to formula (\ref{vibadga3}) can be derived automatically by AD; then one has just to write a computer program that implements the formula of proposition \ref{vibb1} and apply automatic differentiation to the computer program. 
We recall here the basis of AD.
		
		\subsection{Automatic Differentiation}
		
		Consider a function $z=f(u)$ implemented in \texttt{C} or \texttt{C++} by
		\[ \hbox{\texttt{ double f(double u)}} \{...\} \]
		To find an approximation of $z'_u$, one could call in \texttt{C}
		\[ \hbox{\texttt{ double dxdu= (f(u + du)-f(u))/du}} \]
		because
		\[ z'_u=f'(u)=\frac{f(u+\d u)-f(u)}{\d u} + O(|\d u|). \]
		A good precision ought to be reached by choosing $\d u$ small. However arithmetic truncation  limits the accuracy 
		and shows that it is not easy to choose $\d u$ appropriately because beyond a certain threshold, the accuracy of the finite difference formula degenerates due to an almost zero over almost zero ratio. 
		\begin{figure}
			[htbp] 
			\begin{minipage}
				[b]{0.49\textwidth} \centering 
				\includegraphics[width=5cm]{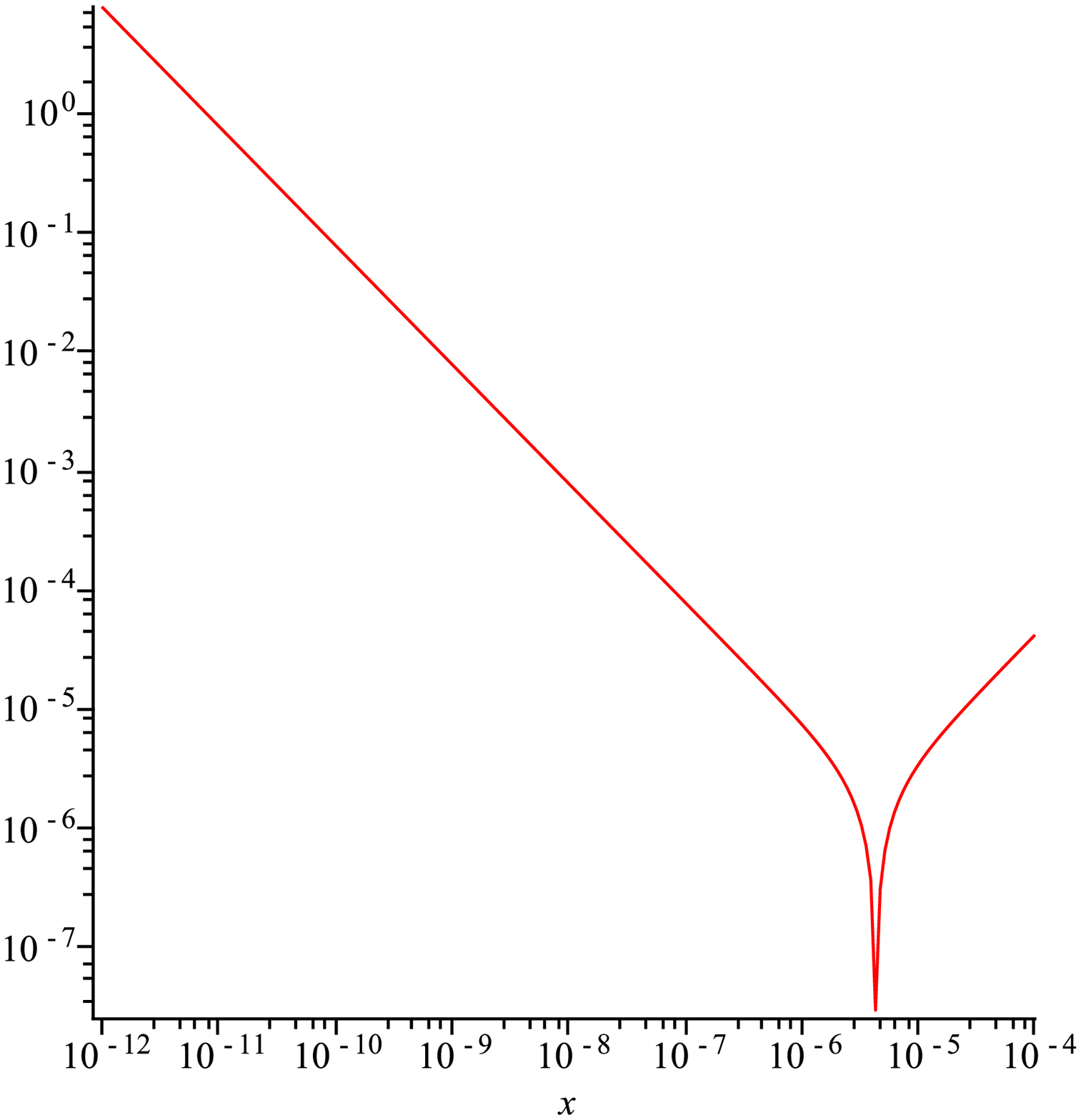} \caption{{\it Precision (log-log plot of $|dzdu-cos(1.)|$ computed with the forward finite difference formula to evaluate $sin'(u)$ at $u=1$.}} \label{figADonea} 
			\end{minipage}
			\hspace{0.3cm} 
			\begin{minipage}
				[b]{0.49\textwidth} \centering 
				\includegraphics[width=5cm]{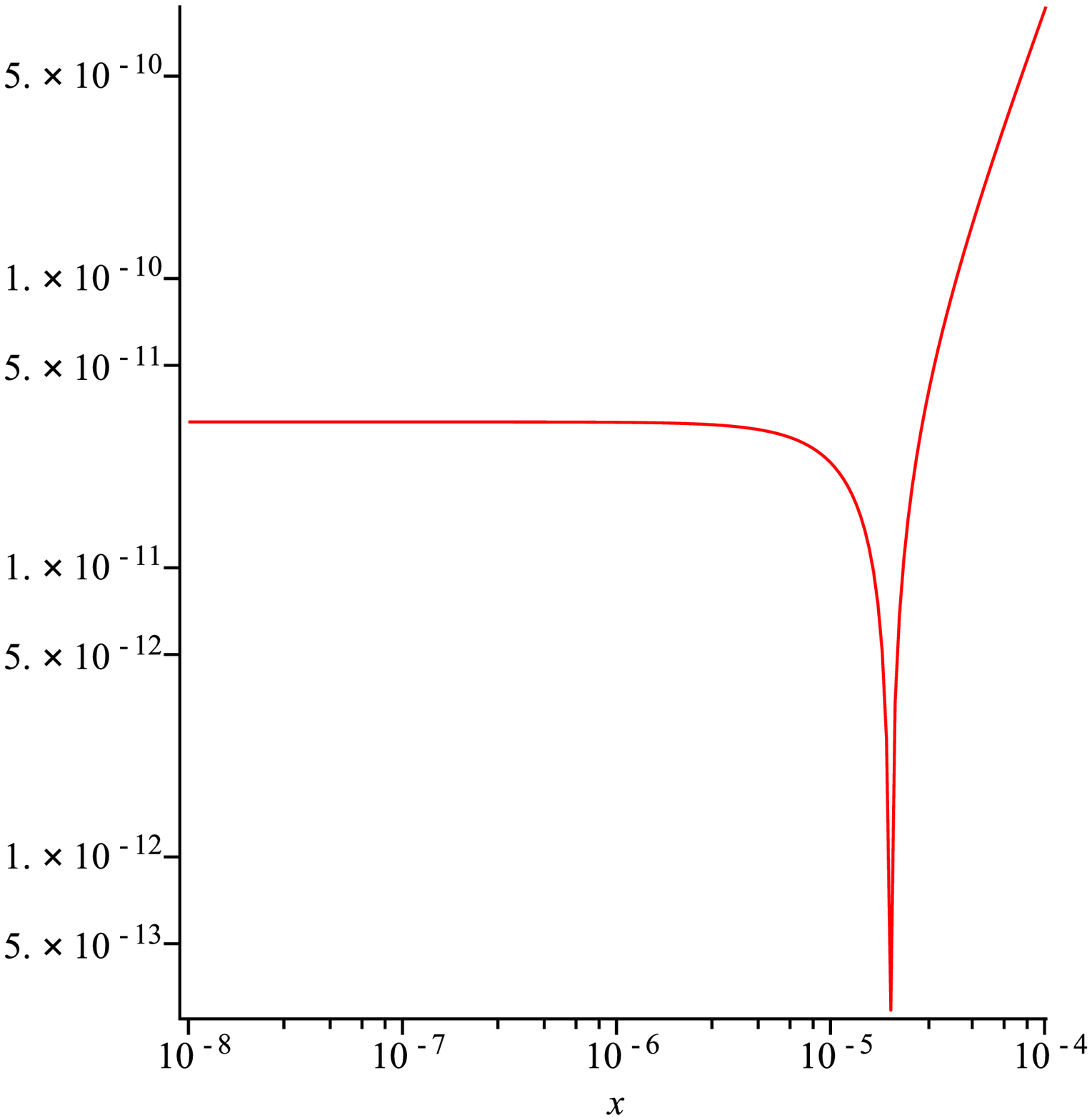} \caption{{\it Same as Fig. \ref{figADonea} but with the finite difference which uses complex increments; both test have been done with Maple-14}} \label{figADoneb} 
			\end{minipage}
		\end{figure}
		As described in Squire et al. \cite{ST98}, one simple remedy is to use complex imaginary increments because
		\[ Re\frac{f(u+\ii\d u)-f(u)}{\ii\d u}= Re\frac{f(u+\ii\d u)}{\ii\d u}=f'(u) - Re f^{'''}(u+\ii\theta\d u)\frac{\d u^2}{6} \]
		leads to $f'(u)=Re[f(u+\ii\d u)/(\ii\d u)]$ where the numerator is no longer the result of a difference of two terms.  
		Indeed tests show that the error does not detoriate when $\d u\to 0$ (figure \ref{figADoneb}). Hence one can choose $\d u=10^{-8}$ to render the last term with a $O(10^{-16})$ accuracy thus obtaining an essentially exact result.
		
		The cost for using this formula is two evaluations of $f()$, and the programming requires to redefine all \texttt{double} as \texttt{std::complex} of the Standard Template Library in C++.
		
		\subsection{ AD in Direct Mode}\label{ADD}
		
		A conceptually better idea is based on the fact that each line of a computer program is differentiable except at switching points of branching statements like \texttt{if} and at zeros of the \texttt{sqrt} functions etc.
		
		Denoting by \texttt{dx} the differential of a variable \texttt{x}, the differential of \texttt{a*b} is \texttt{da*b+a*db}, the differential of \texttt{sin(x)} is \texttt{cos(x)dx}, etc$\ldots$ By operator overloading, this algebra can be built into a C++ class, called \texttt{ddouble} here: 
		\begin{verbatim}
class ddouble { 
    public: double val[2]; 
    ddouble(double a=0, double b=0){ val[1]=b; val[0]=a; } 
    ddouble operator=(const ddouble& a)
      { val[1] = a.val[1]; val[0]=a.val[0]; return *this; } 
    ddouble operator - (const ddouble& a, const ddouble& b)
      { return ddouble(a.val[0] - b.val[0],a.val[1] - b.val[1]); } 
    ddouble operator * (const ddouble& a, const ddouble& b)
      { return ddouble(a.val[0] * b.val[0], a.val[1]*b.val[0] 
                              + a.val[0] * b.val[1]); } 
... }; 
\end{verbatim}
		So all \texttt{ddouble} variables have a 2-array of  data: \texttt{val[0]} contains the value of the variable and \texttt{val[1]} the value of its differential.  
Notice that the constructor of \texttt{ddouble} assigns zero by default to \texttt{val[1]}.
		
		To understand how it works, consider the C++ example of figure \ref{add1} which calls a function $f(u,u_d)=(u-u_d)^2$ for $u=2$ and $u_d=0.1$. Figure \ref{add2} shows the same program where \texttt{double} has been changed to \texttt{ddouble} and the initialization of \texttt{u} implies that its differential is equal to 1. The printing statement displays now the differential of $f$ which is also its derivative with respect to $u$ if all parameters have their differential initialized to 0 except $u$ for which has $du=1$. 
{\small{ 		\begin{figure}[htbp] 			
			\begin{minipage}
				[b]{0.45\linewidth} 
\begin{verbatim}
double f(double u, double u_d)
    { double z = u-u_d; 
      return z*(u-u_d); } 
int main() { 
     double u=2., u_d =0.1; 
     cout << f(u,u_d)<< endl; 
     return 0; 
} 
\end{verbatim}
\caption{A tiny C++ program to compute $(u-u_d)^2$ at $u=2, u_d=0.1$.} \label{add1} 
			\end{minipage}
			 \hskip 0.3cm 		
			\begin{minipage}
				[b]{0.45\linewidth} 
\begin{verbatim}
ddouble f(ddouble u, ddouble u_d)
    { ddouble z = u-u_d; 
      return z*(u-u_d); } 
int main() { 
    ddouble u=ddouble(2.,1.), u_d = 0.1; 
    cout << f(u,u_d).val[1] << endl; 
    return 0; 
} 
\end{verbatim}
\caption{The same program now computes $\frac{d}{d u}(u-u_d)^2$ at $u=2, u_d=0.1$.} \label{add2} 
			\end{minipage}
		\end{figure}
}}
		Writing the class double with all functions and common arithmetic operators is a little tedious but not difficult. An example can be downloaded from \texttt{www.ann.jussieu.fr/pironneau}.

The method can be extended to higher order derivatives easily. For second derivatives, for instance, \texttt{a.val[4]} will store $a$, its differentials with respected to the first and second parameter, $d_1a$, $d_2a$ and the second differential $d_{12}a$ where the two parameters can be the same. The second differential of \texttt{a*b} is
$
a*d_{12}b+ d_1a*d_2b+d_2a*d_1b+b*d_{12}a
$, and so on.

		Notice that $\frac{d f}{d u_d}$ can also be computed by the same program provided the first line in the \texttt{main()} is replaced by \texttt{ddouble u=2., u\_d=ddouble(0.1,1.);}. However if both derivatives $\displaystyle\frac{d f}{d u}, \frac{d f}{d u_d}$ are needed, then, either the program must be run twice or the class ddouble must be modified to handle partial derivatives. In either case the cost of computing $n$ partial derivatives  will be approximately $n$ times that of the original program; the reverse mode does not have this numerical complexity and must be used when, say, $n>5$ if expression templates with traits are used in the direct mode and $n>5$ otherwise \cite{Pir08}.

		\subsection{AD in Reverse Mode}
		
		Consider  finding $F'_{\theta}$ where $(u,\theta)\to F(u,\theta)\in\R$ and $u\in\R^d$ and $\theta\in\R^n$. Assume that $u$ is the solution of a well posed linear system $A u = B \theta + c$.
		
		The direct differentiation mode applied to the C++ program which implements $F$ will solve the linear system $n$ times at the cost of $d^2n$ operations at least.
		
		The mathematical solution by calculus of variations starts with
		\[ F'_{\theta} \d \theta = (
		\partial_{\theta}F) \d \theta + (
		\partial_u F) \d u \hbox{ with } A \d u = B \d \theta, \]
		then introduces $p\in\R^{d}$ solution of $ A^T p=(
		\partial_u F)^T$ and writes
		\[ (
		\partial_u F) \d u = (A^Tp)^T\d u=p^TB\d \theta~\Rightarrow~~ F'_{\theta} \d \theta = (
		\partial_{\theta}F+p^TB)\d \theta. \]
		The linear system for $p$ is solved only once, i.e. performing $O(d^2)$ operations at least. Thus, as the linear system is usually the costliest operation, this second method is the most advantageous when $n$ is large.
		
		A C program only made of assignments can be seen as a triangular linear system for the variables. Loops can be unrolled and seen as assignments and tests, etc. Then, by the above method, the i$^{th}$ line of the program is multiplied by  $p_i$ and $p$ is computed from the last line up; but the biggest difficulty is the book-keeping of the values of the variables, at the time $p$ is computed. 
		
		For instance, for the derivative of \texttt{f=u+ud} with respect to \texttt{ud} with \texttt{u} given by \texttt{\{u=2*ud+4; u=3*u+ud;}\},\texttt{u} in the second line is not the same as \texttt{u} in the third line and the program should be rewritten as \texttt{u1=2*ud+4; u=3*u1+ud;}. Then the system for p is \texttt{p2=1; p1=3*p2;} and the derivative is \texttt{2*p1+p2+1=8}.
				
		\medskip
		
		In this study we have used the library \texttt{adept 1.0} by R.J. Hogan described in Hogan \cite{Hog14}. The nice part of this library is that the programming for the reverse mode is quite similar to the direct mode presented above; all differentiable variables have to be declared as \texttt{ddouble} and the variable with respect to which things are differentiated is indicated at initialization, as above.
		
\subsection{Non-Differentiable Functions}\label{nonDiff}
In finance, non-differentiability is everywhere.  For instance, the second derivative in $K$ of $(x-K)^+$  does not exist at $x=K$ as a function, yet the second derivative of $\int_0^\infty f(x)(x-K)^+ d x$ is $f(K)$. Distribution theory extends the notion of  derivative:  the Heavyside function $H(x)={\bf 1}_{\{x\ge 0\}}$ has the Dirac mass at zero $\delta(x)$ for derivative.

Automatic differentiation can be extended to handle this difficulty to some degree by approximating the Dirac mass at $0$ by the functions $\delta^a(x)$ defined by
\[
\delta^a(x) = \frac1{\sqrt{a\pi}}e^{-\frac{x^2}a}.
\]
Now, suppose $f$ is discontinuous at $x=z$ and smooth elsewhere; then
\[
f(x) = f^+(x)H(x-z)+f^-(x)(1-H(x-z))
\]
hence
\[
f'_z(x) = (f^+)'_z(x)H(x-z) + (f^-)'_z(x)(1-H(x-z)) -(f^+(z)-f^-(z))\delta(x-z)
\]
Unless this last term is added, the computation of the second order sensitivities will not be right.

If in the AD library the ramp function  $x^+$ is defined as $xH(x)$ with its derivative to be $H(x)$, if $H$ is defined with its derivative equal to $\delta^a$ and if in the program which computes the financial asset it is written that
$
(x-K)^+ = \hbox{ramp}(x-K)
$, then the second derivative in $K$ computed by the AD library will be $\delta^a(x-K)$. Moreover, it will also compute
\[
\int_0^\infty f(x)(x-K)^+ d x\approx \frac1N\sum _{i=1}^N f(\xi_i)\delta^a(\xi_i-K)
\]
where $\xi_i$ are the $N$ quadrature points of the integral or the Monte-Carlo points used by the programmer to approximate the integral.

However, this trick does not solve all problems and one must be cautious; for instance writing that $(x-K)^+ = (x-K)H(x-K)$ will not yield the right result.
Moreover, the precision is rather sensitive to the value of $a$.

\begin{remark}
Notice that finite difference (FD) is not plagued by this problem, which means that FD with complex increment is quite a decent method for first order sensitivities. For second order sensitivities the ``very small over very small" problem is still persistent. 
\end{remark}
  
\section{VAD and the Black-Scholes Model}\label{APPLICATION}
		In this section, we implement and test VAD and give a conceptual algorithm that describes the implementation of this method (done automatically).  We focus on indicators which depend on the solution of an SDE, instead of the solution of the SDE itself. Let us take the example of a standard European Call option in the Black-Scholes model.

\subsection{Conceptual algorithm for VAD}\label{BSALGO}
		
\begin{enumerate}
			\item \label{GENE} Generate $M$ simulation paths with time step $h=\frac{T}{n}$ of the underlying asset $X$ and its tangent process $\displaystyle Y=\frac{
			\partial X}{
			\partial \theta}$ with respect to a parameter $\theta$ for $k=0,\dots,n-2$:
			\begin{equation}
				\left\{ 
				\begin{alignedat}
					{2} &\bar X^n_{k+1}= \bar X^n_{k}+rh\bar X^n_{k}+\bar X^n_{k} \sigma\sqrt{h}Z_{k+1},~~\bar X^n_0=X_0,\bar Y^n_0=\frac{
					\partial X_0}{
					\partial \theta}&&\\
					&\bar Y^n_{k+1}= \bar Y^n_{k}+rh\bar Y^n_{k}+\frac{
					\partial}{
					\partial \theta}\left(rh\right)\bar X^n_{k}+\left(\bar Y^n_{k}\sigma\sqrt{h}+\frac{
					\partial}{
					\partial \theta}\left(\sigma\sqrt{h}\right)\bar X^n_{k}\right)Z_{k+1},&&~~.
				\end{alignedat}
				\right. 
			\end{equation}
			\item For each simulation path 
			\begin{enumerate}
				\item Generate $M_Z$ last time steps $( \bar X_T =\bar X^n_{n})$ 
				\begin{equation}
					\bar X^n_{n}= \bar X^n_{n-1}(1+rh+\sigma\sqrt{h}Z_{n}) .
				\end{equation}
					
					\item \label{BSVIB} Compute the first derivative with respect to $\theta$ by Vibrato using the antithetic technique (formula (\ref{vibb1}) with $\sigma(X_t)$ equal $X_t\sigma$)
					\begin{eqnarray} \label{algoVib2}
						\frac{
						\partial V_T}{
						\partial\theta}&=& \frac{
						\partial \mu_{n-1}}{
						\partial \theta}\frac{1}{2}(V_{T_+}- V_{T_-})\frac{Z_{n}}{\bar X^n_{n-1}\sigma \sqrt{h}}
						\cr&&
						+\frac{
						\partial \sigma_{n-1}}{
						\partial \theta}\frac{1}{2}(V_{T_+}-2V_{T_\bullet}+V_{T_-})\frac{Z^2_{n}-1}{\bar X^n_{n-1}\sigma \sqrt{h}} .
					\end{eqnarray}
					With $V_{T_{\pm,\bullet}}=(\bar X_{T_{\pm,\bullet}}-K)^+$,
					\begin{equation}
						\left\{
						\begin{alignedat}
							{3} &\bar X_{T_\pm}=\bar X^n_{n-1}+rh\bar X^n_{n-1}\pm\sigma\bar X^n_{n-1}\sqrt{h}Z_{n}\\
							& \bar X_{T_\bullet}=\bar X^n_{n-1}+rh\bar X^n_{n-1}.
						\end{alignedat}
						\right. 
					\end{equation}
					and 
					\begin{eqnarray}
						&&\frac{
						\partial \mu_{n-1}}{
						\partial \theta}=\bar Y^n_{n-1}(1+rh)+\bar X^n_{n-1}\frac{
						\partial }{
						\partial \theta}(rh)\cr &&\frac{
						\partial \sigma_{n-1}}{
						\partial \theta}=\bar Y^n_{n-1}\sigma\sqrt{h}+X^n_{n-1}\frac{
						\partial}{
						\partial \theta}(\sigma \sqrt{h}) 
					\end{eqnarray}
					If $\theta=T$ or $\theta=r$, we have to add $\displaystyle \frac{
					\partial}{
					\partial \theta}(e^{-rT})V_T$ to the result above.

				\item \label{BSAUTO} Apply an Automatic Differentiation method on the computer program that implements step \ref{algoVib2} to compute the second derivative with respect to $\theta$ at some $\theta^*$. 
			\item \label{BSVIBB} Compute the mean per path i.e. over $M_Z$.
			\end{enumerate}
			\item Compute the mean of the resulting vector (over the $M$ simulation paths) and discount it. 
		\end{enumerate}

\subsection{Greeks}	
The Delta measures the rate of changes in the premium $\E[V(X_T)]$ with respect to changes in the spot price $X_0$. \\
	
The Gamma measures the rate of changes of the Delta with respect to changes in the spot price. Gamma can be important for a Delta-hedging of a portfolio.
\\\\
	The Vanna is the second derivative of the premium with respect to $\sigma$ and $X_0$. The Vanna measures the rate of changes of the Delta with respect to changes in the volatility.
		
\subsection{Numerical Test} \label{NUMTEST}

For the generation of the random numbers, we chose the standard Mersenne-Twister generator available in the version 11 of the C++ STL. We take $M_Z=1$ i.e. we simulate only one last time step per path; for all the test cases except for the European Call contract in the Black-Scholes model. However, for the European Call in a Black-Scholes model, we used a multiple time steps with the Euler scheme with or without a Brownian bridge. \\\\
		The parameters considered in the following numerical experiments are $K=100$, $\sigma=20\%$ and $r=5\%$, $T=1$ year. The initial price of the risky asset price is varying from $1$ to $200$. The Monte Carlo parameters are set to $100,000$ simulation paths, $25$ time steps.
		
\subsubsection{Preliminary Numerical Test}		
 Here, we  focus on the numerical precision of VAD on the Gamma of a standard European Call contract with  constant volatility and drift for which there is an analytical Black Scholes formula. Since Vibrato of Vibrato is similar to Vibrato+AD (VAD) it is pointless to compare the two.
		%

Recall (Proposition \ref{propVAD} \& \ref{propVIBVIB}) that it is equivalent to apply Vibrato to Vibrato or to apply automatic differentiation  to Vibrato. However, the computation times are different and naturally double Vibrato is faster.


We compare the analytical solution to those obtained with VAD but now for each new set of parameters, we reuse the same sample of the random variables.
		
\begin{figure}[ht!] 
\centering 
			\includegraphics[width=0.49\textwidth]{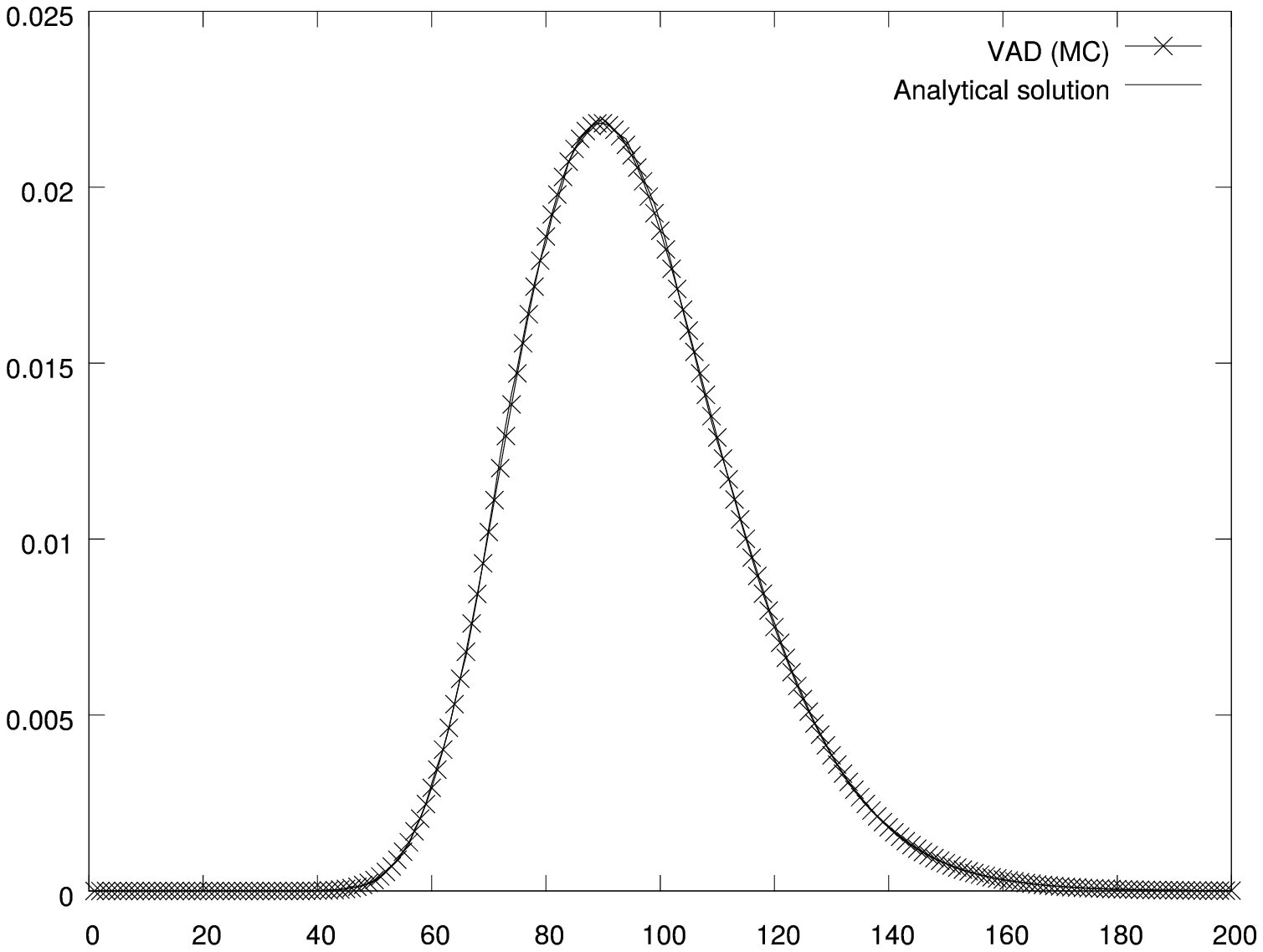}
			\includegraphics[width=0.49\textwidth]{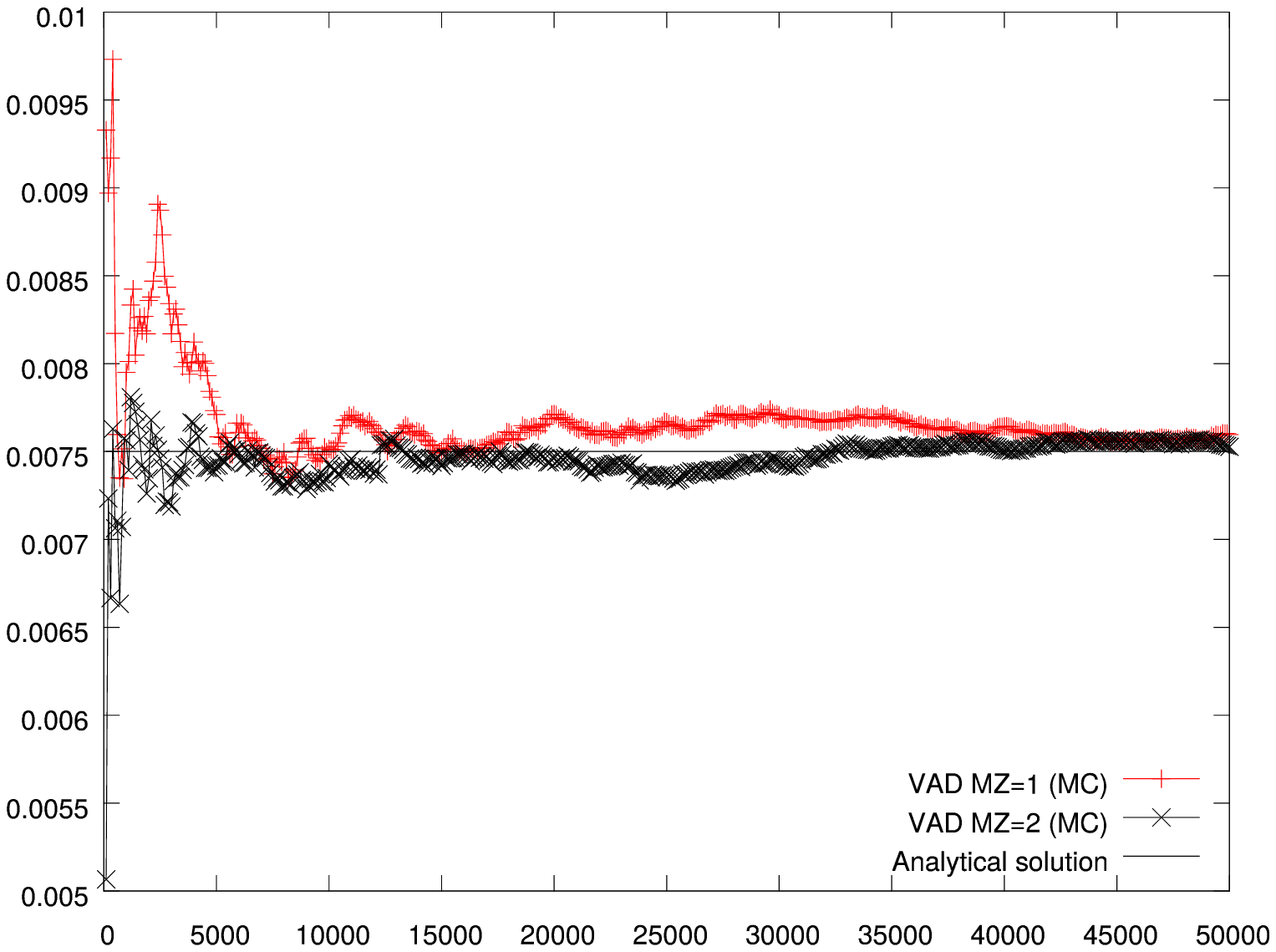}
\caption{\label{gamCal}{\it On the left the Gamma versus Price is displayed when computed by VAD; the analytical exact Gamma is also displayed; both curves overlap. On the right, the convergence history at one point $X_0=120$ is displayed with respect to the number of Monte Carlo samples $M_W$. This is done for two values of $M_Z$ (the number of the final time step), $M_Z=1$ (low curve) and $M_Z=2$ (upper curve).} }
\end{figure}

On figure \ref{gamCal}, the Gammas are compared at $X_0=120$; true value of the Gamma is $\Gamma_0= 0.0075003$. The convergence with respect to the number of paths is also displayed for two values of $M_Z$. The method shows a good precision and fast convergence when the number of paths for the final time step is increased. 

\begin{figure} [ht!] \centering 
			\includegraphics[width=0.49
			\textwidth]{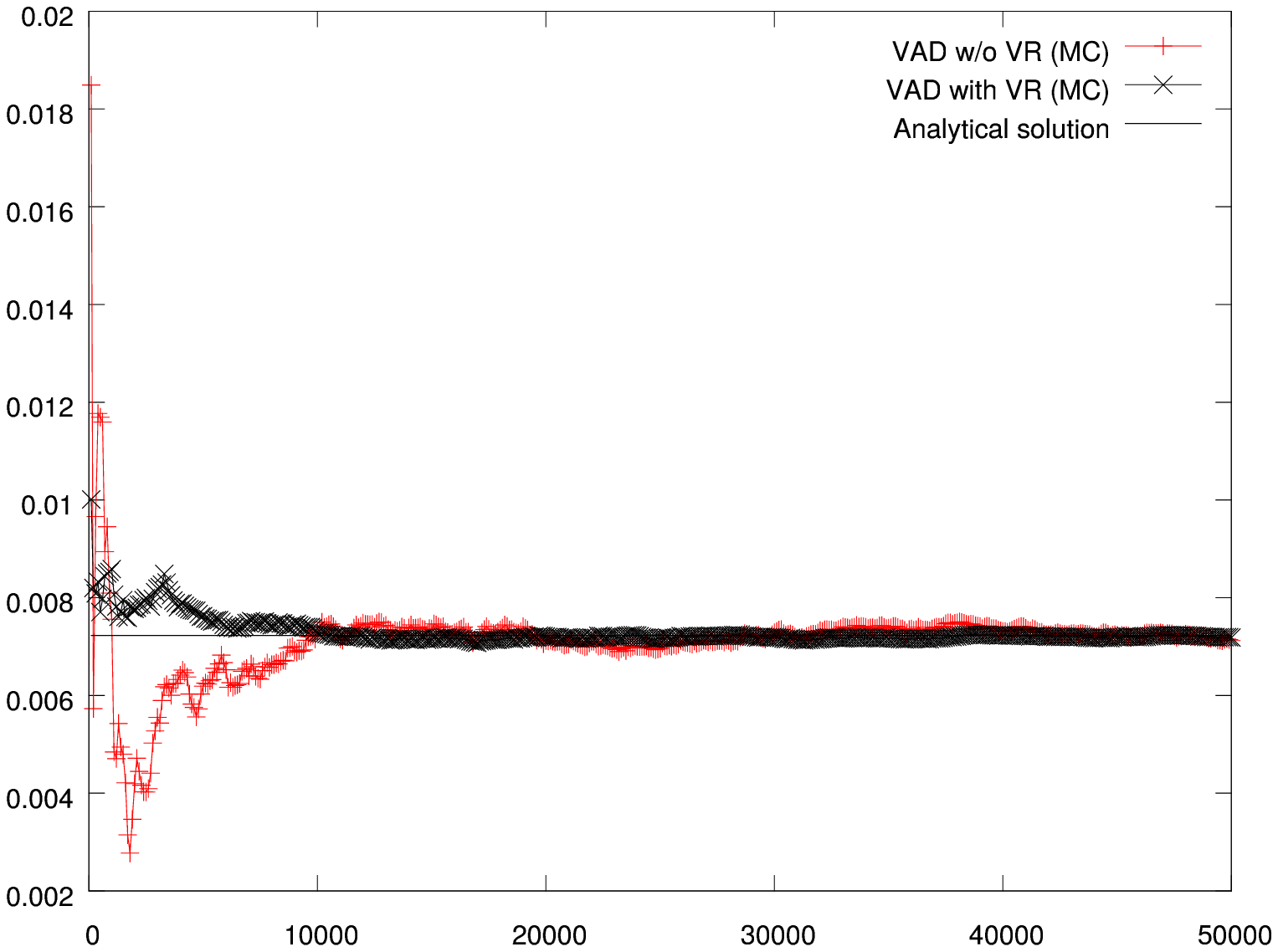} 
			\includegraphics[width=0.49
			\textwidth]{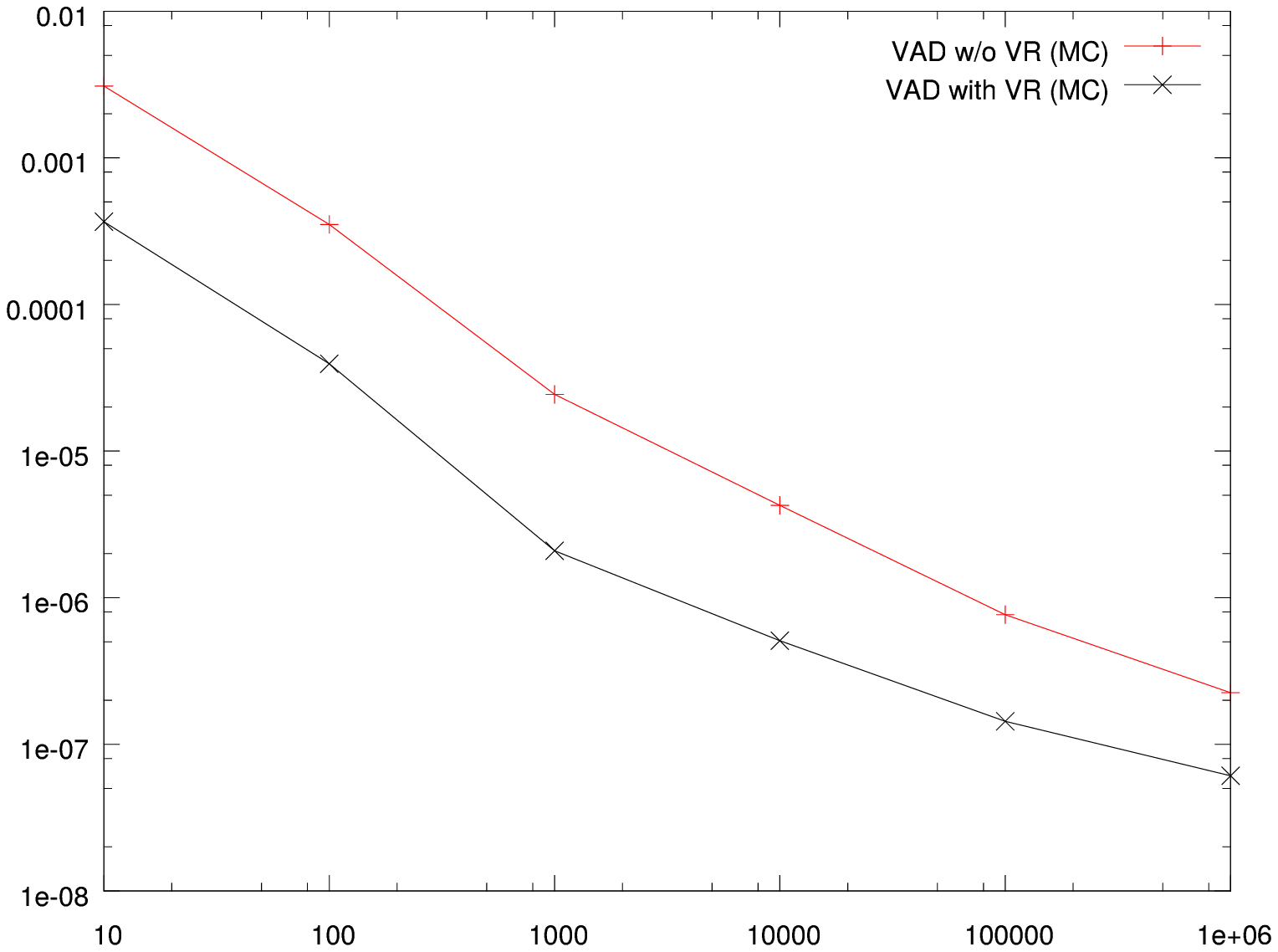} \caption{\label{VRcall}{\it On the left the Gamma versus the number of simulation paths is displayed when computed by VAD with and without the variance reduction method on $Z$, the straight line is the analytical solution at one point $X_0=120$; On the right, the standard error of the two methods versus the number of simulation paths with and without variance reduction.} }
		\end{figure}

The $L^2$-error denoted by $\varepsilon _{L^2}$ is defined by
			\begin{equation}
				\varepsilon_{L^2}=\frac{1}{P}\sum_{i=1}^P(\bar\Gamma ^i-\Gamma _0)^2.
			\end{equation}
On figure \ref{VRcall}, we compare the results with and without variance reduction on Vibrato at the final time step i.e. \!antithetic variables. The convergence history against the number of simulation paths is displayed. Results show that variance reduction is efficient on that test case. The standard error against the number of simulation paths is also displayed. It is clear that a reduction variance is needed. It requires almost ten times the number of simulation paths without the reduction variance technique to obtain the same precision. The Gamma is computed for the same set of parameters as given above. 
		
\begin{figure}[ht!] \centering 
			\includegraphics[width=0.49\textwidth]{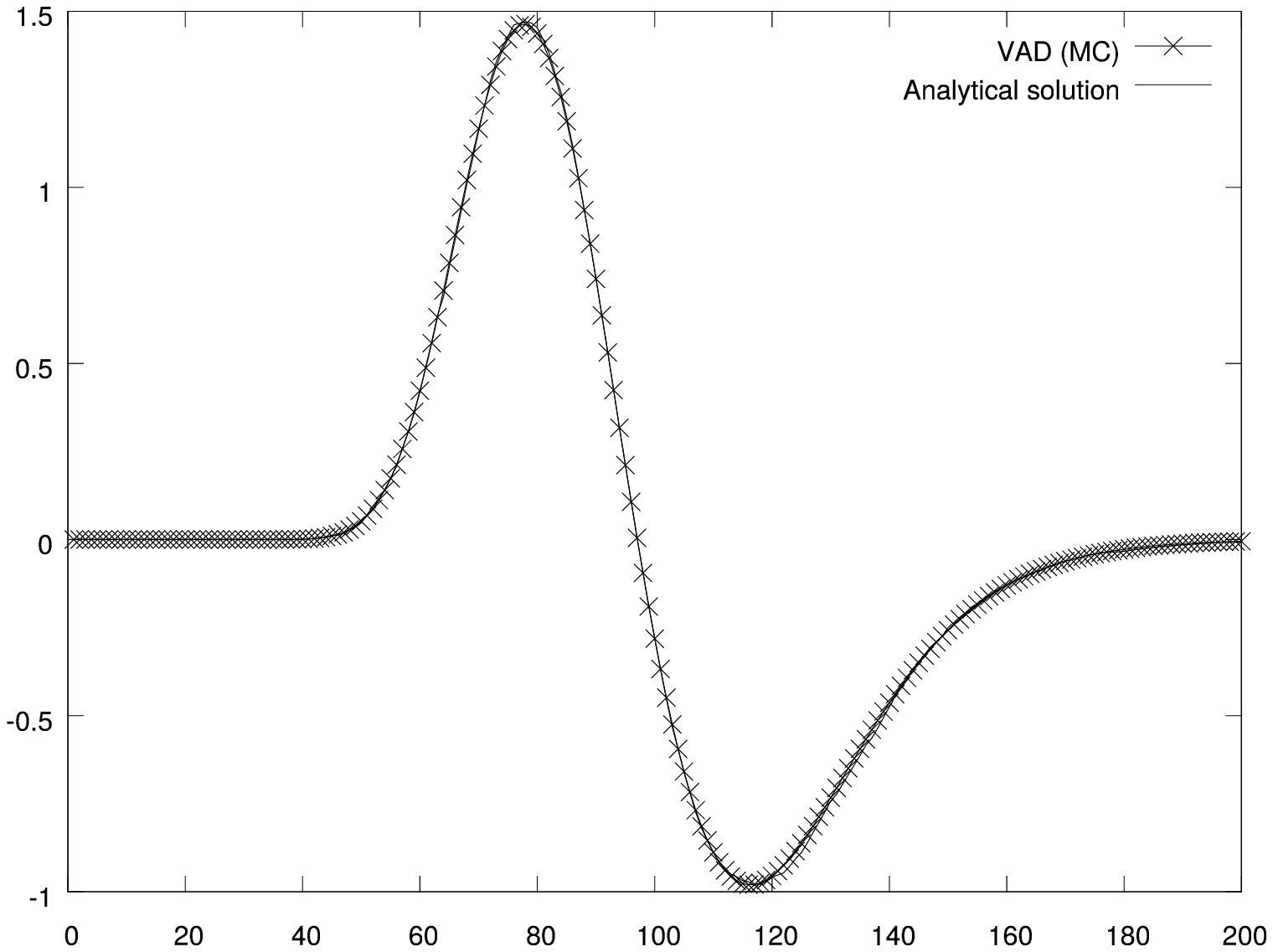} 
			\includegraphics[width=0.49\textwidth]{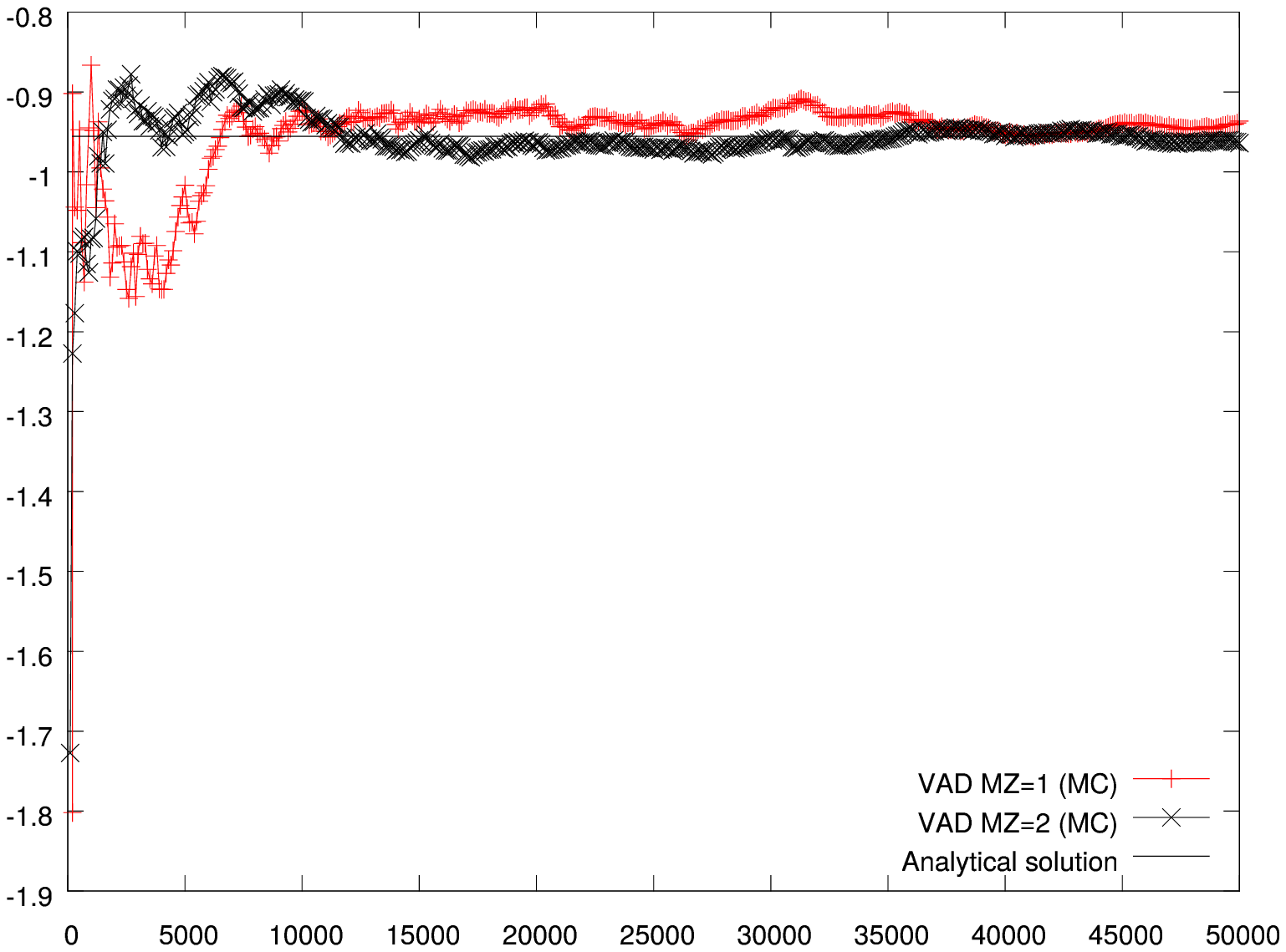}
			\caption{\label{vanCal}{\it On the left the Vanna versus Price is displayed when computed by VAD; the analytical exact Vanna is also displayed; both curves overlap. On the right, the convergence history at one point $X_0=120$ is displayed with respect to the number of Monte Carlo samples $M_W$. This is done for two values of $M_Z$, $M_Z=1$ (lower curve) and $M_Z=2$ (upper curve).} }
		\end{figure}

		On figures \ref{vanCal} 
		we display  the Vanna 
		of an European Call option,  computed with VAD. And again, the convergence with respect to the number of simulation paths is accelerated by more sampling of the final  time step. Note that the Vanna requires double the number of time steps 
%
		%
		
\subsubsection{Third Order Derivatives} 
\begin{figure}
			[htbp!] \centering 
			\includegraphics[width=0.49
			\textwidth]{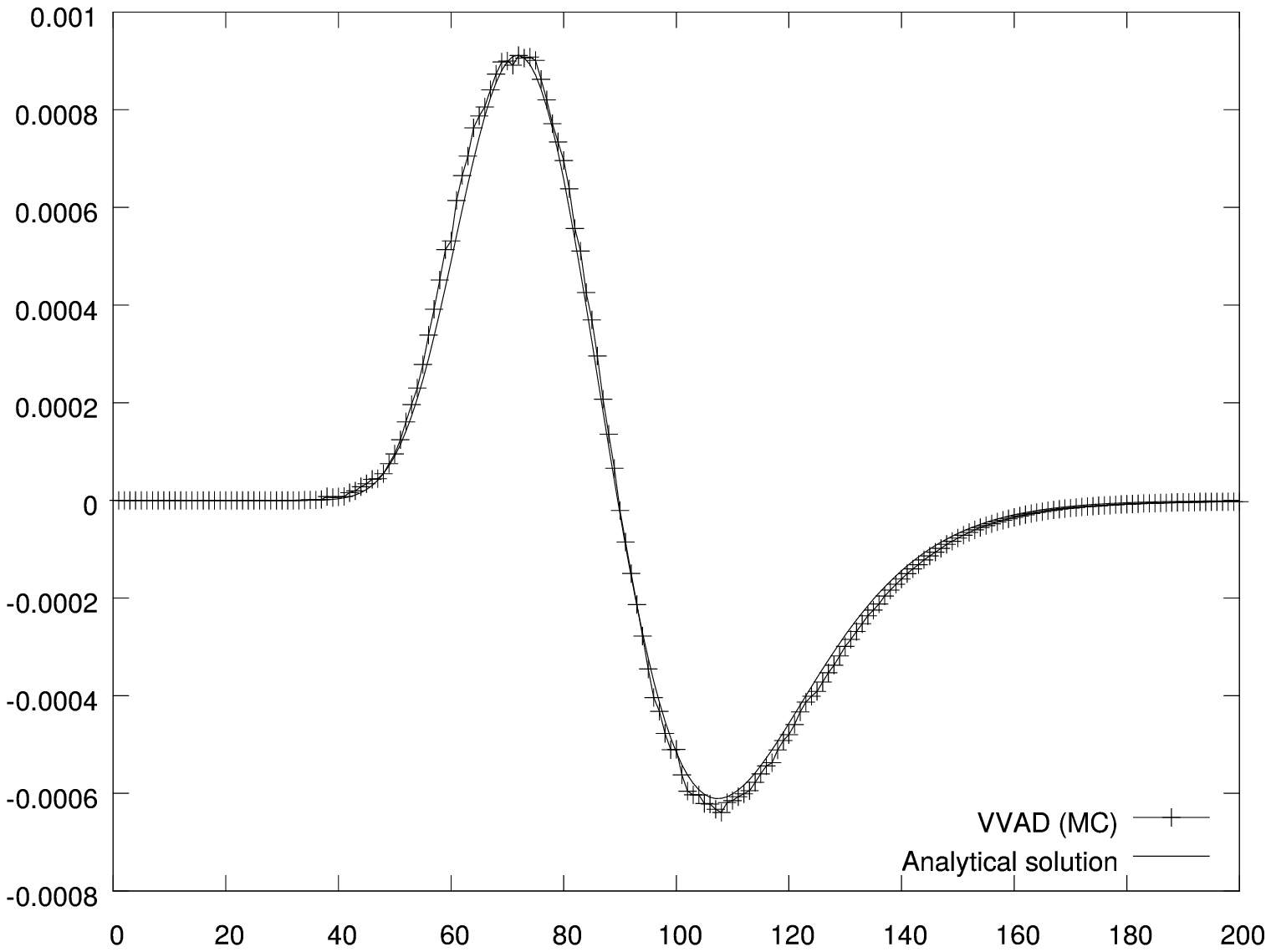} 
			\includegraphics[width=0.49
			\textwidth]{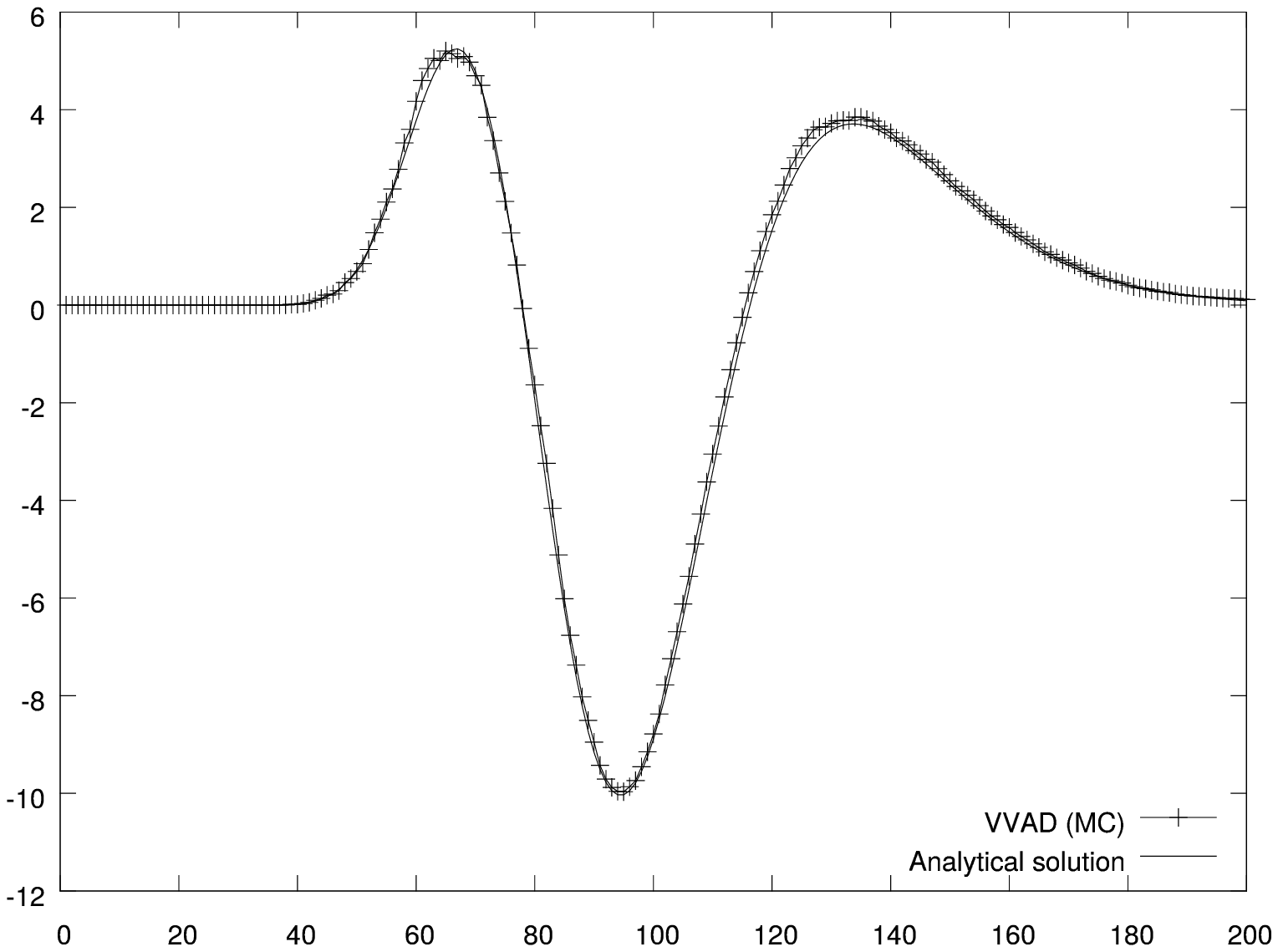} \caption{\label{3rd}{\it On the left $\partial ^3V/\partial {X_0^3}$ versus Price is displayed when computed by VVAD; the analytical exact curve is also displayed; both curves practically overlap. On the right, the same  for the Vanna with respect to changes in interest rate ($\partial ^3V/\partial{X_0}\partial  \sigma \partial r$). } }
		\end{figure}

		For third order derivatives, we compute second derivatives by Vibrato of Vibrato \ref{propVAD} and differentiate by AD (VVAD). The sensitivity of the Gamma with respect to changes in $X_0$ is  $\partial ^3V/\partial {X_0^3}$.  The sensitivity of the Vanna with respect to changes in the interest rate is  $\partial ^3V/\partial{X_0}\partial  \sigma \partial r$. The parameters of the European Call are the same but the Monte Carlo path number is $1,000,000$ and $50$ time steps for the discretization. The results are displayed on figure \ref{3rd}. The convergence is slow; we could not eliminate the small difference between the analytical solution and the approximation by increasing the number of paths. 
		
		\subsubsection{Ramp Function and High Order Derivatives}

As mentioned in Section \ref{nonDiff}, it is possible to handle the non-differentiability of the function $(x-K)^+$ at $x=K$ by using  distribution theory and program  the ramp function explicitly with a second derivative equal  to an approximate Dirac function at K. We illustrate this technique with a standard European Call option in the Black-Scholes model. We computed the Gamma and the sixth derivative with respect to $X_0$. For the first derivative, the parameter $a$ does not play an important role but, as we  evaluate higher derivatives, the choice of the parameter $a$ becomes crucial for the quality of a good approximation and it requires more points to catch the Dirac approximation with small $a$.  Currently the choice of $a$ is experimental.

\begin{figure} [ht!] \centering 
			\includegraphics[width=0.49
			\textwidth]{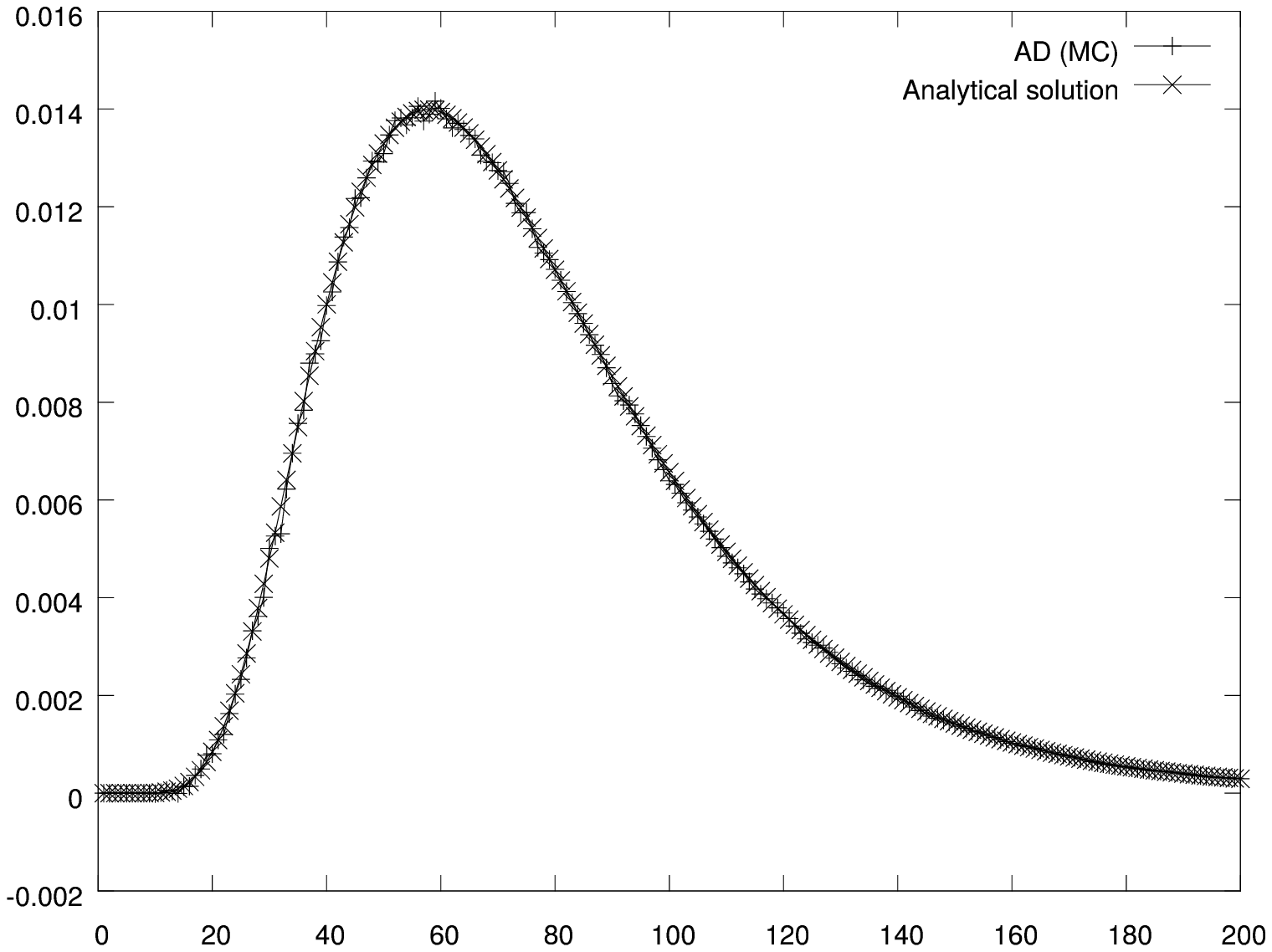} 
			\includegraphics[width=0.49
			\textwidth]{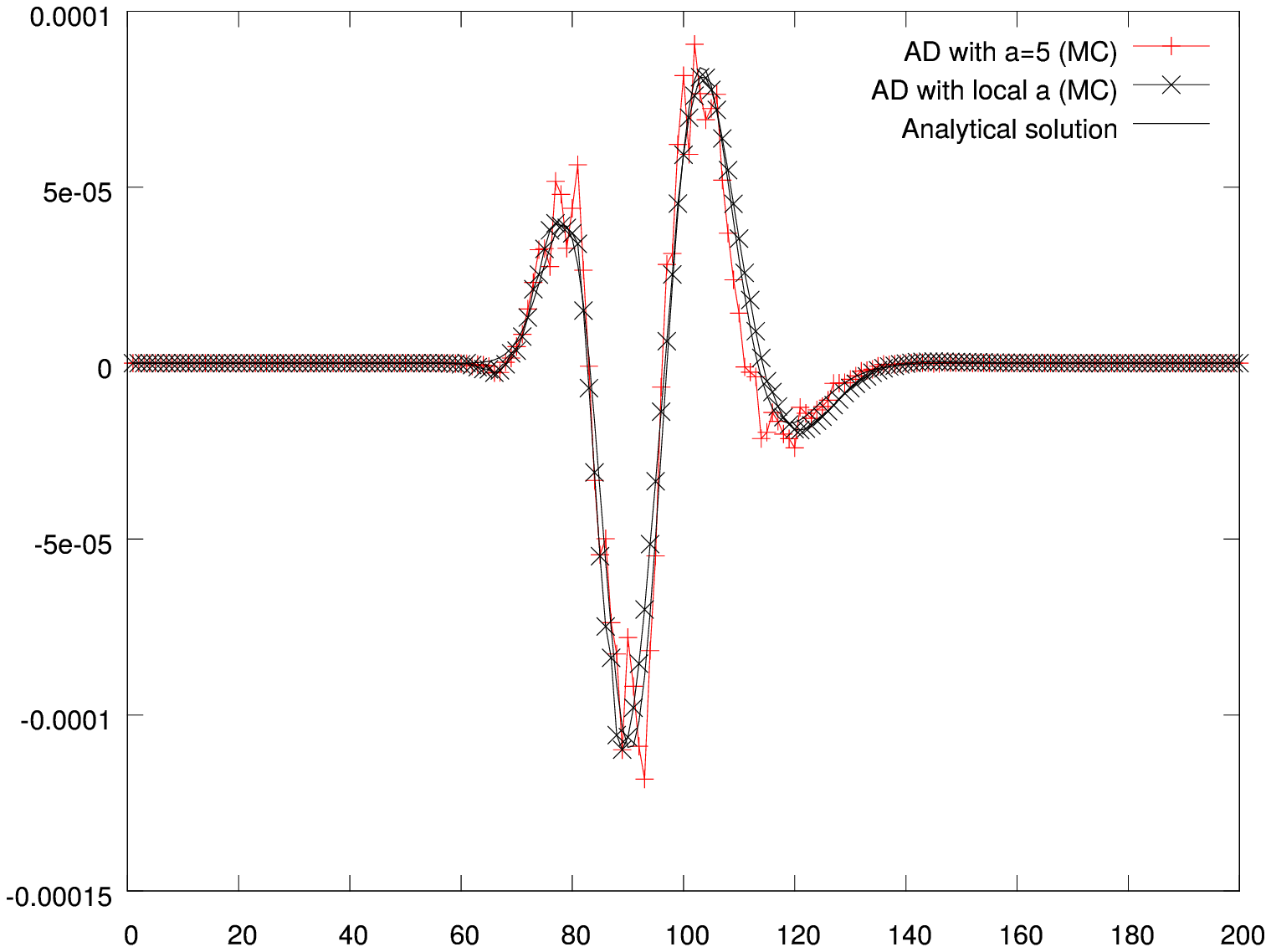} \caption{\label{NONdif}\it On the left the Gamma versus Price is displayed when computed by AD with the ramp function (with $a=1$); the analytical exact Gamma is also displayed; both curves overlap. On the right, the sixth derivative with respect to the parameter $X_0$ is displayed when computed via the same method; the analytical solution is also displayed. We computed the approximation with local parameter $a$ and with $a=5$.  }
		\end{figure}

	We took the same parameters as previously for the standard European Call option but the maturity for the Gamma now set at $T=5$ years and $T=0.2$ year for the sixth derivative with respect to  $X_0$. The initial asset price varies from $1$ to $200$. The Monte Carlo parameters are also set to $100,000$ simulation paths and $25$ time steps. The results are displayed on figure \ref{NONdif}. 
		
For the Gamma, the curves are overlapping but for the sixth derivative with respect to the parameter $X_0$, we cannot take a constant parameter $a$ anymore. When we choose locally adapted parameter $a$, the curves are practically overlapping.

		\subsection{Baskets}
		A Basket option is a multidimensional derivative security whose payoff depends on the value of a weighted sum of several risky underlying assets. 
		
As before, $X_t$ is given by  (\ref{1e}).		
But now $(W_t)_{t\in[0,T]}$ is a $d$-dimensional {\it correlated} Brownian motion with $ {\mathbb E}[dW^i_tdW^j_t]=\rho_{i,j}dt$. 

To simplify the presentation, we assume that $r$ and $\sigma_i$ are real constants and the payoff is given by 
		\begin{equation}
			\textbf{V}_T =e^{-rT} \textbf{E}[( \sum _{i=1}^d \omega_i {X_i}_T -K ) ^+] 
		\end{equation}
		where $(\omega _i)_{i=1,\dots,d}$ are positive weights with $\sum_{i=1}^d\omega _i = 1$. Here, we choose to compare three different methods. The reference values coming from an approximated moment-matching dynamics (Levy \cite{Lev92} and in Brigo et al. \cite{BMRS02}), VAD and second order finite difference (FD). 
		
\subsubsection{Algorithm to compute the Gamma of a Basket option} 
		We make use of the fact that $r$ and $\sigma$ are constant.
		\begin{enumerate}
			\item Generate $M$ simulation paths using a one time step for the Euler scheme. 
			\begin{equation}
				\nonumber \bar {X^i}_{T\pm }= {X^i}_{T_\bullet}\exp{\left(- \frac{1}{2}\sum_{j=1} ^d |\Sigma ^{ij}|^2 T \pm \sum _{j=1}^d \Sigma ^{ij}\sqrt{T} Z_j\right)},~~i=1,\dots,d,  
			\end{equation}
			with $ X_{T_\bullet} =  {X}_{0}\exp{(rT)}$,
			where $Z$ denotes an $\mathcal{N}(0;I_{d})$ random vector. 

			\item For each simulation path, with $C=\Sigma \Sigma ^\textbf{T}$, compute (Vibrato)
				\begin{eqnarray}
					\Delta&=& \left(\frac{
					\partial \mu}{
					\partial {X_i}_0}\right)^T\frac{1}{2\sqrt h}( V_{T_+}-V_{T_-})C^{-T}Z
					\cr&+&\frac{1}{4h}( V_{T_+}-2V_{T_\bullet}+V_{T_-})\frac{
					\partial \Sigma}{
					\partial {X_i}_0}:C^{-T}(ZZ^T-I_d)C^{-1}
				\end{eqnarray}
				with $V_{T_.}=(\omega\cdot\bar X_{T_.}-K)^+$
			\item Compute the mean of the resulting vector and discount the result.
			\item Apply Automatic Differentiation to what precedes.
		\end{enumerate}

		\subsubsection{Numerical Test}
		
		In this numerical test $d=7$ and the underlying asset prices are: 
		\begin{equation}
			{X_0}^T = (1840,1160,3120,4330.71,9659.78,14843.24,10045.40).
		\end{equation}
		The volatility vector is: 
		\begin{equation}
			{\sigma} ^T = (0.146,0.1925,0.1712,0.1679,0.1688,0.2192,0.2068) .
		\end{equation}
		The correlation matrix is 
		\begin{equation}
			\begin{pmatrix}
				1.0&0.9477&0.8494&0.8548&0.8719&0.6169&0.7886 \\
				0.9477&1.0&0.7558&0.7919&0.8209&0.6277&0.7354 \\
				0.8494&0.7558&1.0&0.9820&0.9505&0.6131&0.9303 \\
				0.8548&0.7919&0.9820&1.0&0.9378&0.6400&0.8902 \\
				0.8719&0.8209&0.9505&0.9378&1.0&0.6417&0.8424 \\
				0.6169&0.6277&0.6131&0.6400&0.6417&1.0&0.5927 \\
				0.7886&0.7354&0.9303&0.8902&0.8424&0.5927&1.0 
			\end{pmatrix}.
		\end{equation}
		The number of Monte Carlo paths varies from 1 to $10^6$ with only one time step for the time integration. Errors are calculated with reference to a solution computed by approximate moment matching.
		
\begin{figure}[ht!] 
		\begin{minipage}[b]{0.45\linewidth}
			\centering 
			\includegraphics[width=1.1
			\textwidth]{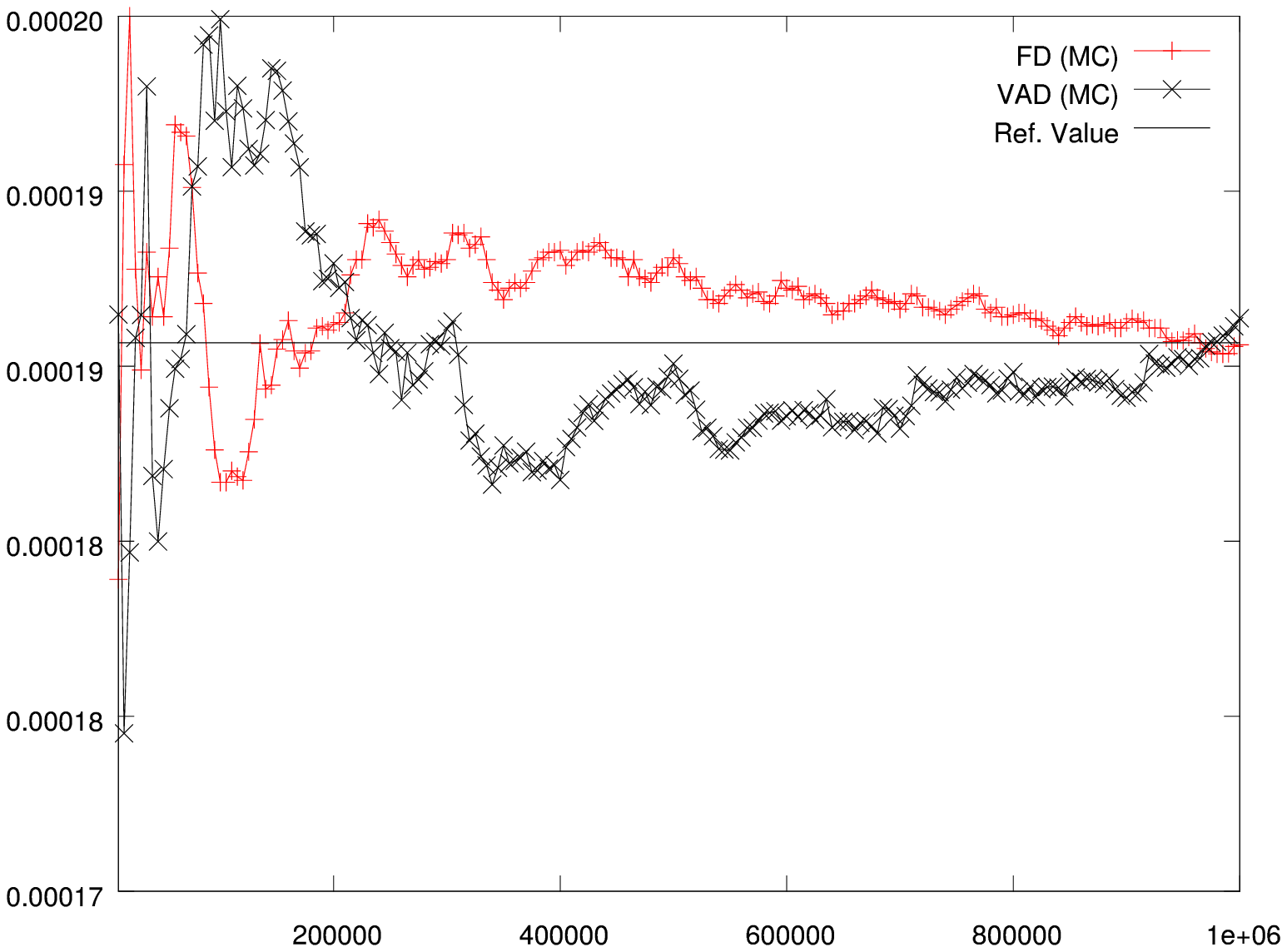} 
			\caption{\label{4dConv}{\it d=4.}} 
		\end{minipage}
%
		\begin{minipage}[b]{0.45\linewidth}
	\centering
			\includegraphics[width=1.1
			\textwidth]{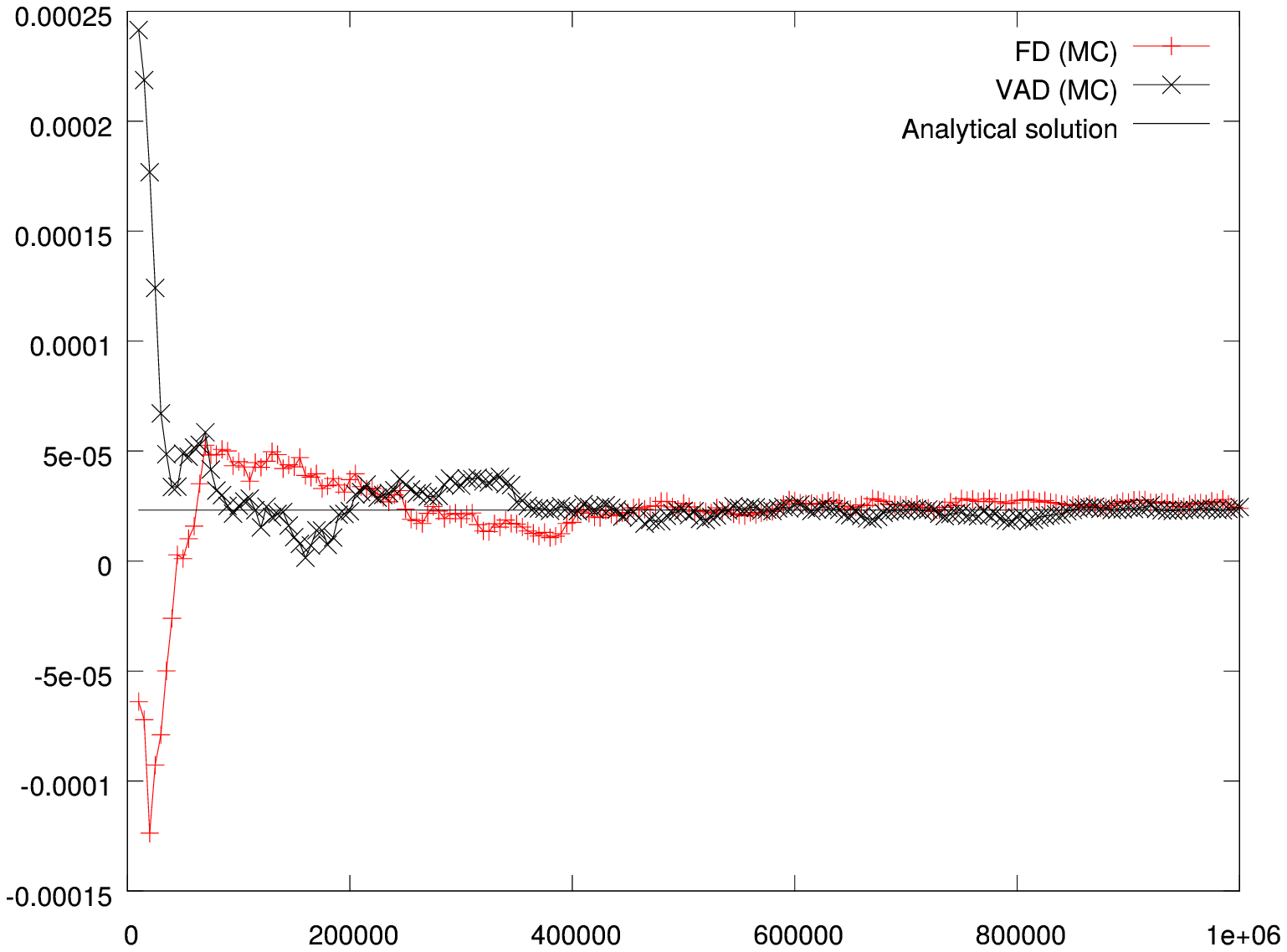}
			\caption{\label{7dConv}{\it d=7.}} 
		\end{minipage}
		{\it 
		
		Convergence of the computation of the Gamma of a Basket option when $d=4$ and $7$ via Vibrato plus Automatic Differentiation on Monte Carlo and via Finite differences, versus the number of simulation paths. The parameters are for $T=0.1$.}
		\end{figure}

		On figures \ref{4dConv} and \ref{7dConv}, the plot of convergence for the computation of the Gamma of a Basket made of the first 4 and 7 assets are displayed versus the number of simulation paths Vibrato plus AD (direct mode) and for Finite differences applied to a brute force Monte Carlo algorithm. The convergence speed of these methods is almost the same (with a slight advantage for the Finite difference). 
		
		Table \ref{7dTab} displays results for a Basket with the $7$ assets, in addition the table \ref{7dTime} displays the CPU time for Vibrato plus AD (direct mode); the finite difference method is one third more expensive. Again, the method is very accurate.


		\section{American Option}\label{AMERICAN}
		
		Recall that an American option is like a European option which can be exercised at any time before maturity. The value $V_t$ of an American option requires the best exercise strategy. Let $\varphi$ be the payoff, then 
		\begin{equation}
			{V}_t:=\underset{\tau \in {\cal T}_{t}}{\text{ess\,sup\,}} \textbf{E}[e^{-r(\tau-t)}\varphi(X_\tau)\mid X_t ] 
		\end{equation}
		where $ {\cal T}_{t}$ denotes the set of $[t,T]$-valued stopping times (with respect to the (augmented) filtration of the process $(X_s)_{s\in[0,T]}$).
		
				Consider a time grid $0<t_1<\dots< t_n=T$ with time step $h$, i.e. $t_k=k h$. To discretize the problem we begin by assuming that the option can be exercised only at $t_k,~k=0,..,n$ ; its value is defined recursively by 		
\begin{equation} \label{ls}
			\left\{
			\begin{alignedat}
				{3} &\bar{V}_{t_n} = e^{-r T} \varphi(\bar X_{T}) \\
				&\bar{V}_{t_k} = \max_{0\leq k \leq n-1}{\left(e^{-rt_k}\varphi( \bar X_{t_k} ), \mathbb{E}[\bar V_{t_{k+1}} \mid \bar X_{t_k}] \right)}, 
			\end{alignedat}
			\right. 
\end{equation}
		\subsection{Longstaff-Schwartz Algorithm }
Following Longstaff et al. \cite{LS01} let the continuation value $C_{t_k}=\mathbb{E}[e^{-r h }\bar V_{t_{k+1}} \mid {\bar X}_{t_k}]$ as $X$ is a Markov process. The holder of the contract exercises only if the payoff at $t_k$ is higher than the continuation value $C_{t_k}$. The continuation value is approximated by a linear combination of a finite set of  $R$ real basis functions:
		\begin{equation}
			C_{{k}} \simeq \sum_{i=1}^R \alpha _{{k},i}\psi _{{k},i}(\bar X_{{t_k}}).
		\end{equation}
 Typically, the $(\alpha _{{k},i})_{i=1,\dots,R}$ are computed by least squares,
		\begin{equation}
			\min_{\alpha}\left\{\mathbb{E}\left[\left( \mathbb{E}[e^{-r{h} }\bar V_{t_{k+1}} \mid {\bar X}_{{t_k}}]- \sum_{i=1}^R \alpha _{{k},i}\psi _{{k},i}(\bar X_{{t_k}})\right)^2 \right]\right\}.
		\end{equation}
		This leads to a Gram linear system 
		\begin{equation}\label{linsys}
			\sum_{j=1}^R \alpha _{{k},i}\mathbf{Gram}\left\{\psi_{{k},i}(\bar X_{t_k}), \psi _{{k},j}(\bar X_{t_k})\right\} = \mathbb{E}[ \mathbb{E}[e^{-r{h} }V_{{k+1}} \mid X_{t_k}]\psi_{{k},i}(\bar X_{t_k}) ],~~i=1,\ldots,R.
		\end{equation}
		\begin{remark}Once the optimal stopping time is known, the differentiation with respect to $\theta$ of (\ref{ls}) can be done as for a European contract. The dependency of the $\tau ^*$ on $\theta$ is neglected; arguably this dependency is second order but this point needs to be validated.
		\end{remark}
		Hence, the following algorithm is proposed.
		
		\subsection{Algorithm to compute the Gamma of an American option} 
		\begin{enumerate}
			\item Generate $M$ simulation paths of an Euler scheme with $n$ time steps of size $h=\frac{T}{n}$.
			\item Compute the terminal value of each simulation path 
			\begin{equation}
				V_{T} = (K-{\bar X_{T}})^+ 
			\end{equation}
			\item Compute the Gamma of the terminal condition using (\ref{algoVib2}) in section (\ref{BSALGO}) for each simulation path.
			
			\item Iterate from $ n-1$ to $1$ and perform the following at the $k$-th time step. 
			\begin{enumerate}
				\item Solve the Gram linear system (\ref{linsys}).
				\item Calculate the continuation value of each path. 
				\begin{equation}
					C_{k+1}(\bar X_{t_k})=\sum^R_{i=1}\alpha_{k,i}\psi_i(\bar X_k^n).
				\end{equation}
				\item Compute the Gamma by differentiating the Vibrato formula from the time step $k-1$ with respect to $X_{0}$ 
				\begin{eqnarray}
					\tilde \Gamma _{k}&=&\frac{1}{N}\sum _{i=1}^N \frac{
					\partial}{ 
					\partial X_0}\left( \bar Y^n_{k-1} \left(1+rh\right)\frac{1}{2}(\tilde V_{{k_+}}^i-\tilde V_{{k_-}}^i)\frac{Z_{k}^{i}}{ X_0\sigma \sqrt{h}}\right. \\
					&&+\left. \bar Y^n_{k-1}\sigma \sqrt{h}\frac{1}{2}(\tilde V_{k_+}^i-2\tilde V_{k_\bullet}^i+\tilde V_{k_-}^i)\frac{(Z^{i}_k)^2-1}{\bar X_0\sigma \sqrt{h}} \right).
				\end{eqnarray}
\item For $i=1,\dots,M$ 
				\begin{equation}
					\left\{
					\begin{alignedat}
						{3} &V_{k}^i = \tilde V_{k}^i,~~\Gamma_{k}^i =\tilde \Gamma _{k}^i ~~ && \mbox{if }\tilde V_{k}^i \geq C_{k+1}(\bar X_{k}^{n,i}),\\
						&V_{k}^i = e^{-rh}V_{k+1}^i,~~\Gamma_{k}^i = e^{-rh}\Gamma_{k+1}^i ~~ && \mbox{otherwise}
					\end{alignedat}
					\right. 
				\end{equation}
				with $\tilde V_{k+1}=(K-\bar X_{k+1}^n)^+$ and 
				\begin{equation}
					\left\{
					\begin{alignedat}
						{3} &\bar X_{{k_\pm}}= \bar X_{k-1}+rh \bar X_{k-1}\pm\sigma \bar X_{k-1}\sqrt{h}Z_{k}\\
						& \bar X_{{k_\bullet }}= \bar X_{k-1}+rh \bar X_{k-1}. 
					\end{alignedat}
					\right. 
				\end{equation}
			\end{enumerate}
			
\item Compute the mean of the vector $V$ and $\Gamma$.
		\end{enumerate}
		\begin{remark}
			The differentiation with respect to $X_0$ is implemented by automatic differentiation of the computer program. 
		\end{remark}
					
		\subsubsection{Numerical Test}
		
		We consider the following value : $\sigma=20\%$ or $\sigma=40\%$, $X_0$ varying from 36 to 44, $T=1$ or $T=2$ year, $K=40$ and $r=6\%$. The Monte Carlo parameters are: $50,000$ simulation paths and $50$ time steps for the time grid. The basis in the Longstaff-Scharwtz algorithm is $(x^n)_{n=0,1,2}$.
		
\begin{figure}
			[ht!] \centering 
			\includegraphics[width=0.49
			\textwidth]{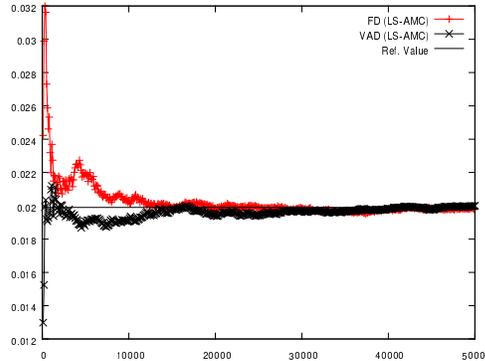}
			
			\caption{\label{amc}Convergence of the Gamma of an American option via Vibrato plus Automatic Differentiation on the Longstaff-Schwartz algorithm and via Finite Difference, versus the number of simulation paths. The parameters are $\sigma=40\%$ and $X_0=40$.
		} 
		\end{figure}

		We compare with the solution of the Black-Scholes partial differential equation discretized by an implicit Euler scheme in time, finite element in space and semi-smooth Newton for the inequalities  \cite{AP05}. A second order finite Difference approximation is used to compute the Gamma. A large number of grid points are used to make it a reference solution. The parameters of the method are $10,000$ and $50$ time steps per year. Convergence history for Longstaff Schwartz plus Vibrato plus AD is shown on figure \ref{amc} with respect to the number of Monte Carlo paths (Finite Difference on Monte Carlo is also displayed). 
		
		On figure \ref{amc}, we display the history of convergence for the approximation of the Gamma of an American Put option versus the number of simulation paths for Vibrato plus Automatic differentiation and for Finite Difference applied to the American Monte Carlo, the straight line is the reference value computed by PDE+ semi-smooth Newton. The convergence is  faster for VAD than with second order Finite Difference (the perturbation parameter is taken as $1\%$ of the underlying asset price).
		
		On table \ref{amctab}, the results are shown for different set of parameters taken from Longstaff et al. \cite{LS01}. The method provides a good precision when  variance reduction (\ref{AMCvar}) is used, for the different parameters, except when the underlying asset price is low with a small volatility. As for the computation time, the method is faster than Finite Difference applied to the American Monte Carlo  which requires three evaluations of the pricing function whereas VAD is equivalent to two evaluations (in direct mode).

		\section{Second Derivatives of a Stochastic Volatility Model}\label{HESTON}
		
		
The Heston model \cite{Hes93} describes the evolution of an underlying asset $(X_t)_{t\in[0,T]}$ with a stochastic volatility $(\mathcal{V}_t)_{t\in[0,T]}$: 
		\begin{eqnarray}
			\label{bs2d} dX_t & = &r X_t dt+ \sqrt{\mathcal{V}_t} X_t dW^1_t, \cr d\mathcal{V}_t &=& \kappa(\eta - \mathcal{V}_t)dt + \xi \sqrt{\mathcal{V}_t}dW_t^2,~~t\in[0,T];~~\mathcal{V}_0,X_0\hbox{ given}. 
		\end{eqnarray}
		 Here $\xi$ is the volatility of the volatility, $\eta$ denotes the long-run mean of $\mathcal{V}_t$ and $\kappa$ the mean reversion velocity. The standard Brownian process $(W_t^1)_{t\in[0,T]}$ and $(W_t^2)_{t\in[0,T]}$ are correlated:  ${\mathbb E}[dW^1_tW^2_t]=\rho dt$, $\rho \in(-1,1)$. If $2\kappa \eta > \xi ^2$, it can be shown that $\mathcal{V}_t>0$ for every $t\! \in [0,T]$. We consider the evaluation of a standard European Call  with payoff 
		\begin{equation}
			{V}_T={\mathbb E}[(X_T-K)^+].
		\end{equation}
		
		\subsection{Algorithm to Compute second derivatives in the Heston Model} 
		To compute the Gamma  by Vibrato method for the first derivative coupled to automatic differentiation for the second derivative one must do the following: 
		\begin{enumerate}
			\item Generate $M$ simulation paths for the underlying asset price $(\bar X,\bar{\mathcal{V}})$ and its tangent process $(\bar Y,\bar{\mathcal{U}}) = \frac{
			\partial (\bar X,\bar{ \mathcal{V}})}{
			\partial X_0}$ using an Euler scheme with $n$ time steps of size $h=\frac{T}{n}$, 
			\begin{equation}
				\left\{ 
				\begin{alignedat}
					{2} &\bar X^n_{{k+1}}= \bar X^n_{{k}} + r h \bar X^n_{{k}} + \sqrt{ \bar{ \mathcal{V}}^n_{{k}} } \bar X^n_{{k}} \sqrt{ h} \tilde Z^1_{k+1},&~~\bar X^n_{0}=X_0, \\
					&\bar Y^n_{k+1} = \bar Y^n_{k} + r h \bar Y^n_{k} + \sqrt{ \bar{ \mathcal{V}}^n_{k} } \bar Y^n_{k} \sqrt{ h} \tilde Z^1_{k+1},&~~\bar Y^n_{0}=1, \\
					&\bar{\mathcal{V}}^n_{{k+1}}=\bar {\mathcal{V}}^n_{{k}}+\kappa (\eta - \bar {\mathcal{V}}^n_{{k}} )h+\xi \sqrt{\bar {\mathcal{V}}^n_{{k}} } \sqrt{h} \tilde Z^2_{k+1},&~~\bar {\mathcal{V}}^n_{0}=\mathcal{V}_0
				\end{alignedat}
				\right. 
			\end{equation}
			with 
			\begin{equation}
				\begin{pmatrix}
					\tilde Z^1 \\
					\tilde Z^2 
				\end{pmatrix}
				= 
				\begin{pmatrix}
					1 & 0 \\
					\rho & \sqrt{1-\rho^2} 
				\end{pmatrix}
				\begin{pmatrix}
					Z^1 \\
					Z^2 
				\end{pmatrix}
			\end{equation}
			where $(Z^1_k,Z^2_k)_{1\leq k \leq n}$ denotes a sequence of $\mathcal{N}(0;I_2)$-distributed random variables. 
			\item For each simulation path 
			\begin{enumerate}
				\item Compute the payoff 
				\begin{equation}
					V_{T}=(\bar X_n^n- K )^+.
				\end{equation}
				
				\item \label{HESAL} Compute the Delta using Vibrato at maturity with the $n-1$ time steps and the following formula 
				\begin{eqnarray}					
					&&
					\bar\Delta^n=\bar Y^n_{n-1}\left(1+rh\right)\frac{1}{2}(V_{T_+}-V_{T_-})\frac{Z^{1}_{n}}{\bar X^n_{n-1}\sqrt{\bar{\mathcal{V}}^n_{n-1}} \sqrt{h}}\\
					&&+ \bar Y^n_{n-1}\sqrt{\bar{\mathcal{V}}^n_{n-1}}\sqrt{h}\frac{1}{2}(V_{T_+}-2V_{T_\bullet}+V_{T_-})\frac{Z^{1^2}_{n}-1}{\bar X^n_{n-1}\sqrt{\bar{\mathcal{V}}^n_{n-1}}\sqrt{h}} 
				\end{eqnarray}
				with 
				\begin{equation}
					\left\{
					\begin{alignedat}
						{1} &\bar X_{T_\pm}= \bar X^n_{n-1} + r h \bar X^n_{n-1} \pm \sqrt{ \bar{ \mathcal{V}}^n_{n-1} } \bar X^n_{n-1} \sqrt{h} \tilde Z^1_{n},\\
						& \bar X_{T_\bullet}= \bar X^n_{n-1}+rh \bar X^n_{n-1}. 
					\end{alignedat}
					\right. 
				\end{equation}
				\item Apply an Automatic Differentiation method on step (\ref{HESAL}) to compute the Gamma. 
			\end{enumerate}
			\item Compute the mean of the result and discount it. 
		\end{enumerate}
		
		\subsubsection{Numerical Test}
		
		We have taken the following values: the underlying asset price $X_0\in[60,130]$, the strike is $K=90$, the risk-free rate $r=0.135\%$ and the maturity o is $T=1$. 
		
		The initial volatility is $\mathcal{V}_0=2.8087\%$, the volatility of volatility is $\xi=1\%$, the mean reversion is $\kappa=2.931465$ and the long-run mean is $\nu=0.101$. The correlation between the two standard Brownian motions is $\rho=50\%$.
		
		The number of Monte Carlo path is $500,000$ with $100$ time steps each.
		
\begin{figure}
			[htbp] \centering 
			\includegraphics[width=0.49
			\textwidth]{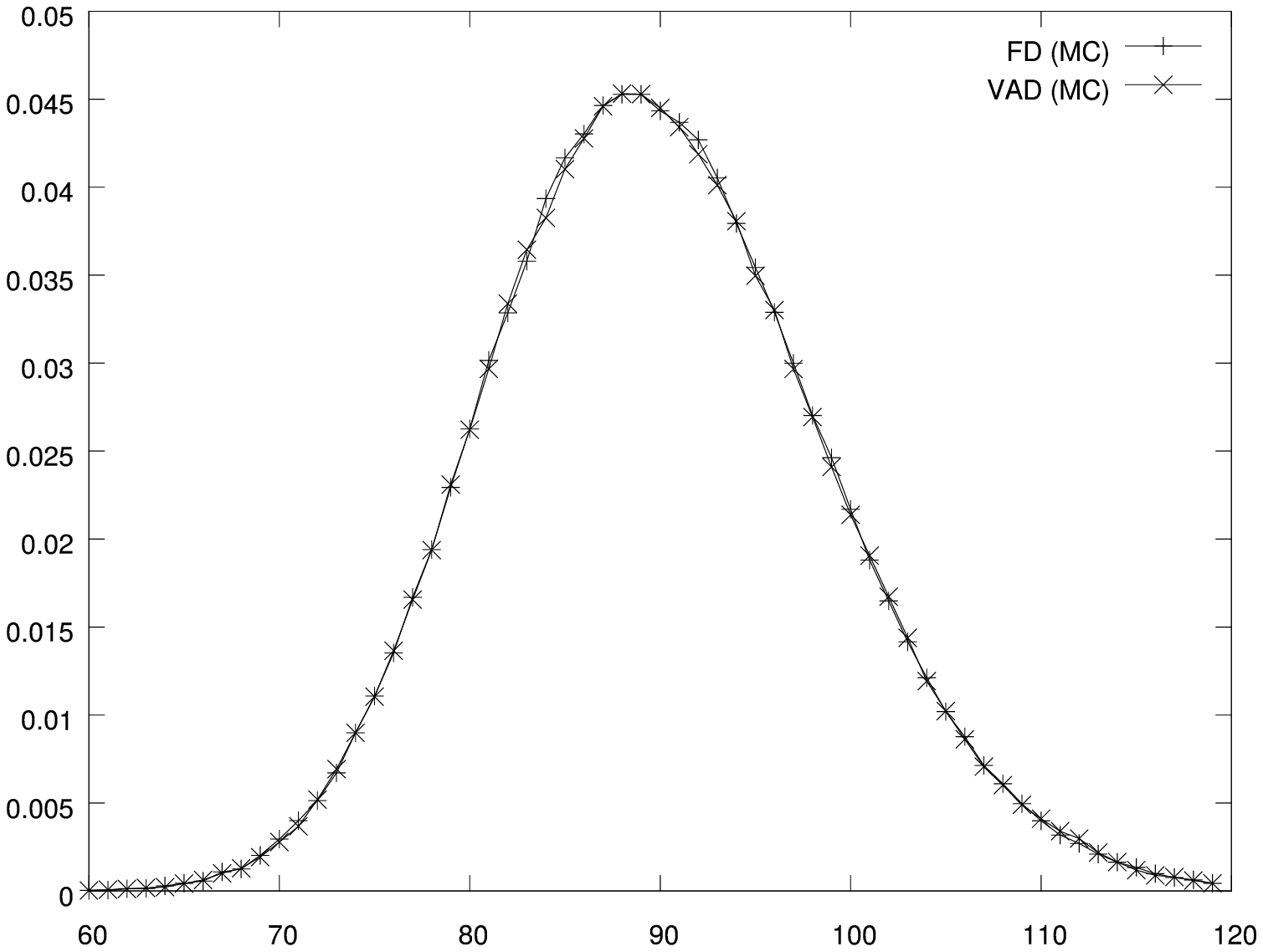} 
			\includegraphics[width=0.49
			\textwidth]{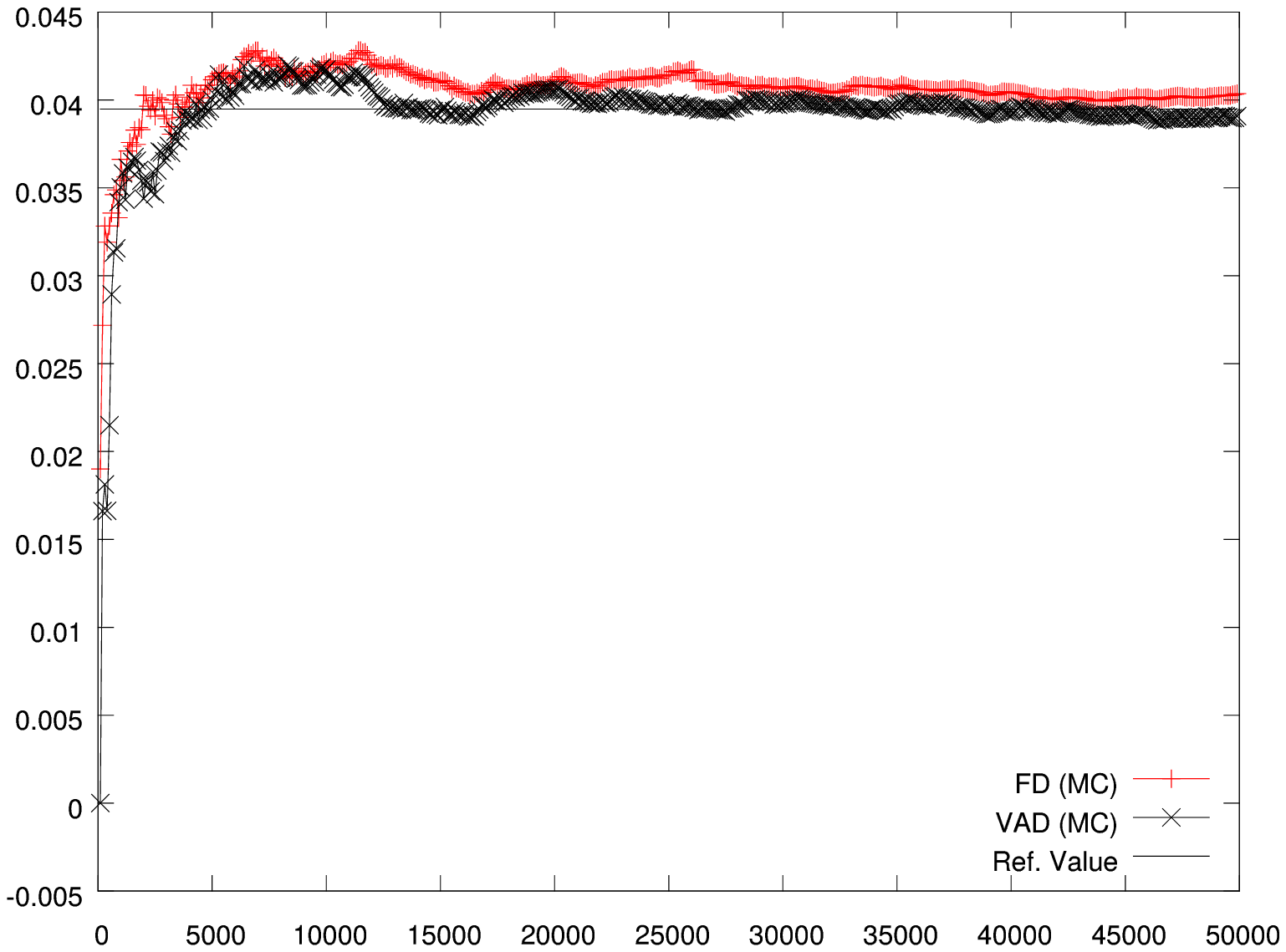} \caption{\label{HESGAM} {\it On the left the Gamma versus Price is displayed when computed by VAD; the approximated Gamma via Finite Difference is also displayed; both curves overlap. On the right, the convergence history at one point $(X_0,\mathcal{V}_0)=(85,2.8087)$ is displayed with respect to the number of Monte Carlo samples.}} 
		\end{figure}

\begin{figure}
			[htbp] \centering 
			\includegraphics[width=0.49
			\textwidth]{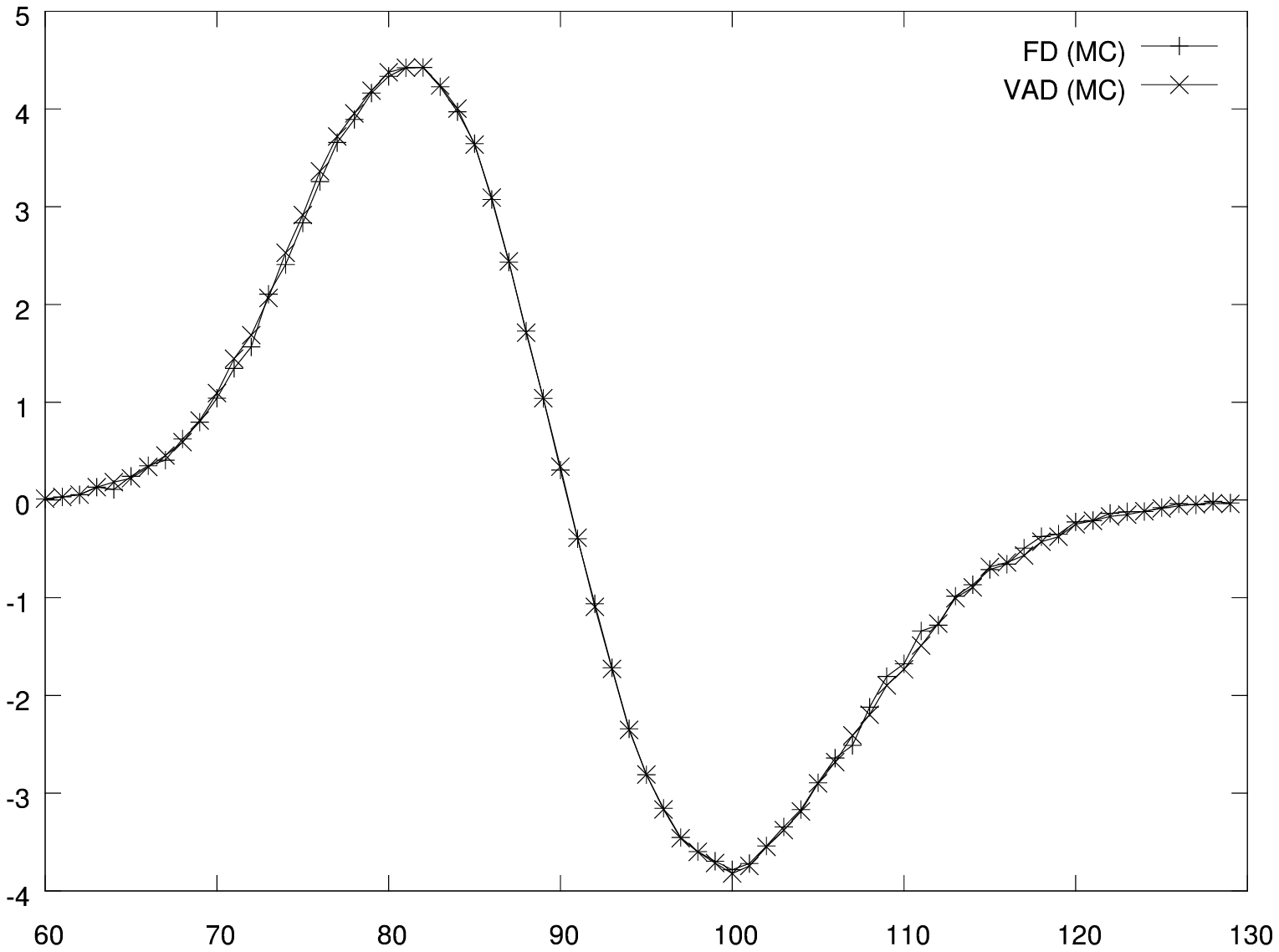} 
			\includegraphics[width=0.49
			\textwidth]{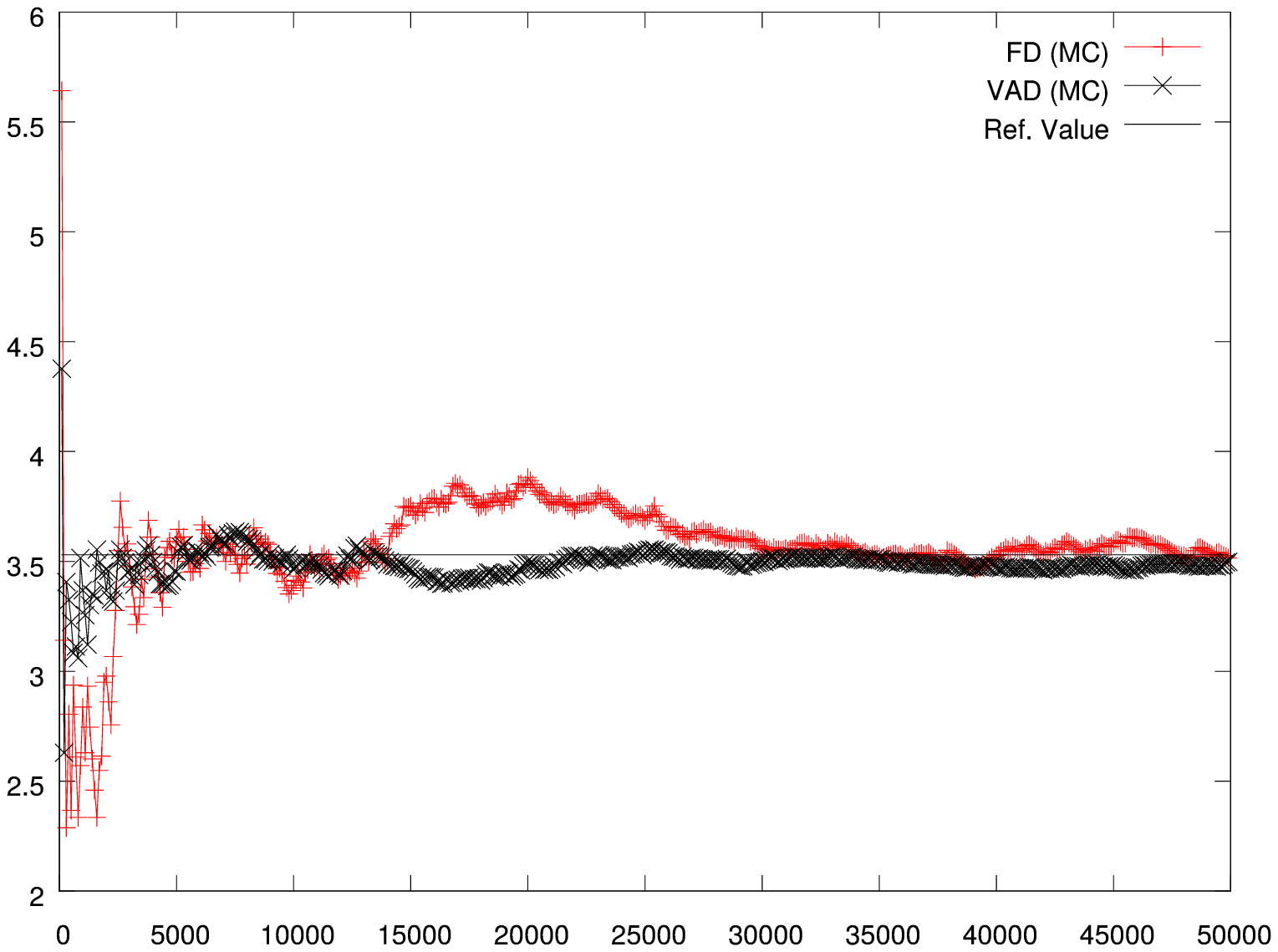}

			\caption{\label{HESVAN}{\it On the left the Vanna versus Price is displayed when computed by VAD; the approximated Vanna via Finite Difference is also displayed; both curves overlap. On the right, the convergence history at one point $(X_0,\mathcal{V}_0)=(85,2.8087)$ is displayed with respect to the number of Monte Carlo samples.}} 
		\end{figure}

		The results are displayed on figures \ref{HESGAM}, \ref{HESVAN}.
		
		On figure \ref{HESGAM} we compare the results obtained by Vibrato plus Automatic Differentiation (direct mode), with second order Finite Difference method applied to a standard Monte Carlo simulation. 
On figures \ref{HESVAN} 
we display  the  Vanna
of an European Call option in the Heston model, and again, the convergence with respect to the number of simulation paths. As for the Gamma, the method is quite precise.
provides a good precision for the approximation of the Vomma and the Vanna. Both are computed at one point $(X_0,\mathcal{V}_0)=(85,2.8087)$ with the same set of parameters as given above. 
		\begin{figure}
			[ht!] \centering 
			\includegraphics[width=0.49
			\textwidth]{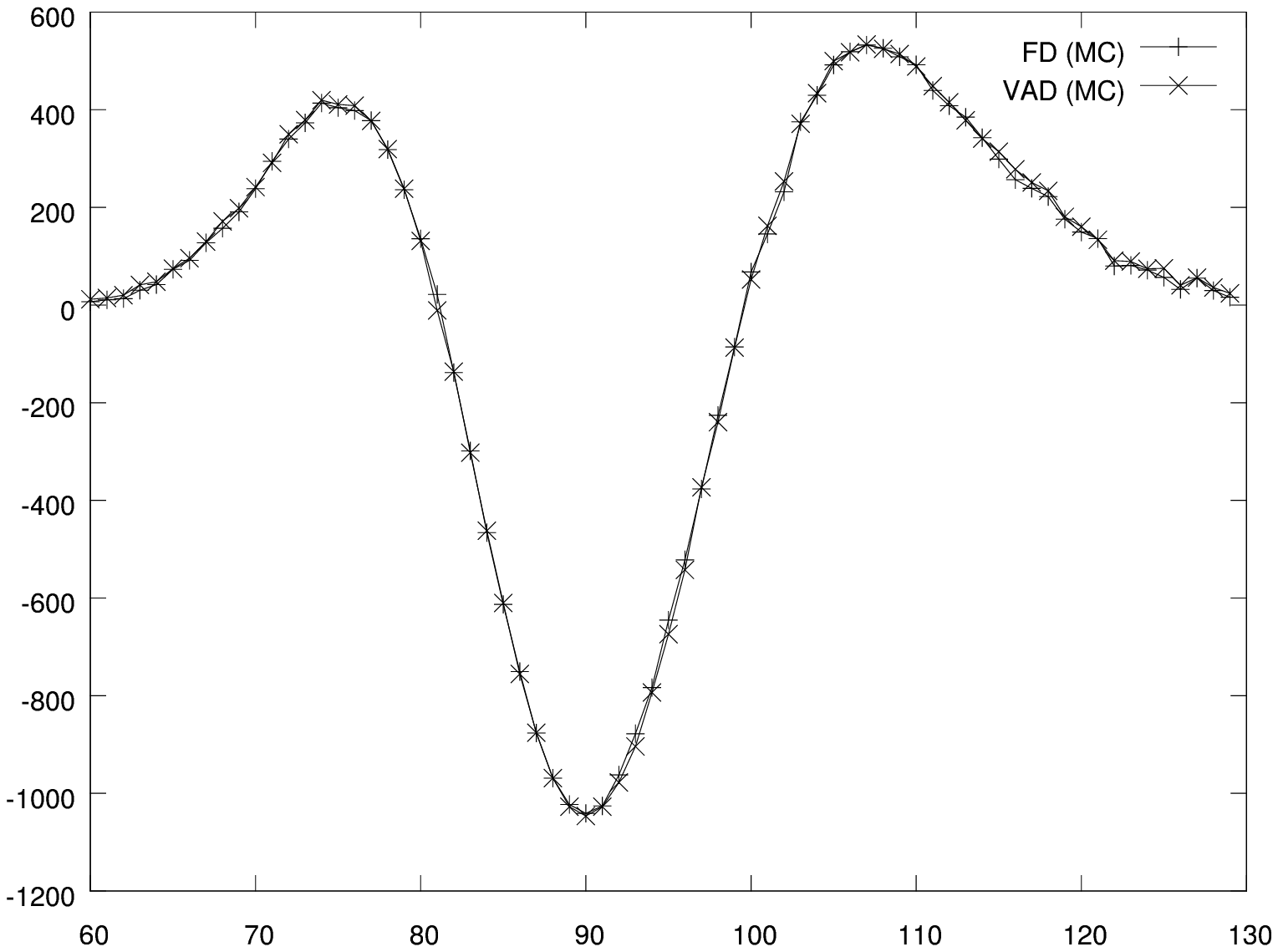} 
			\includegraphics[width=0.49
			\textwidth]{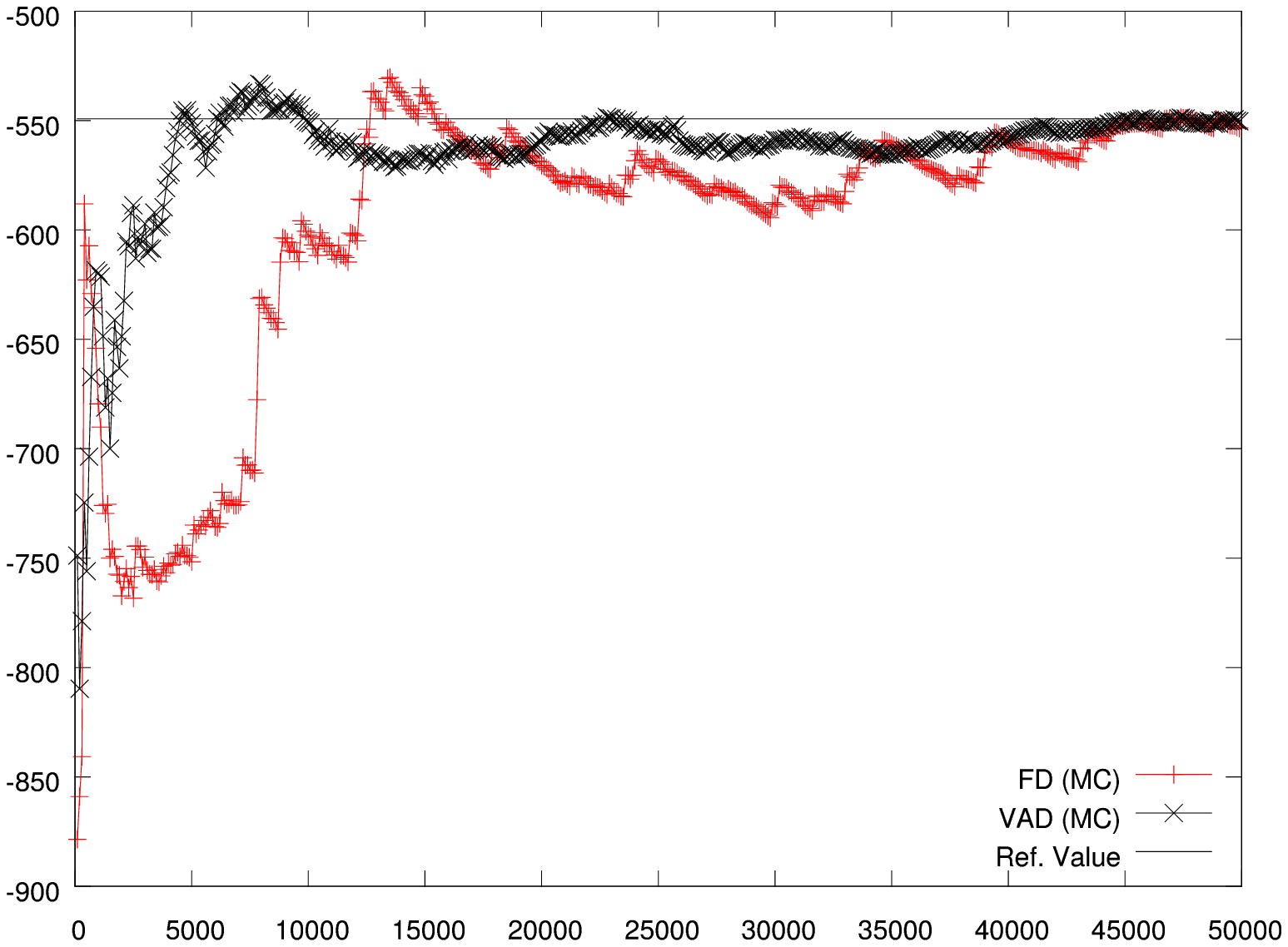}

			\caption{\label{HESVOM}{\it On the left the Vomma versus Price is displayed when computed by VAD; the approximated Vomma via Finite Difference is also displayed; both curves overlap. On the right, the convergence history at one point $(X_0,\mathcal{V}_0)=(85,2.8087)$ is displayed with respect to the number of Monte Carlo samples.}} 
		\end{figure}
		The computation by VAD is 30\% faster for the Gamma compared with the Vanna.
		In the case of the Vomma and the Gamma, VAD is $30\%$ faster. For the Vanna Finite difference requires four times the evaluation of the pricing function so VAD is twice times faster.
		
		
		\section{Vibrato plus Reverse AD (VRAD)} \label{REVERSEAD}
		
If several greeks are requested at once then it is better to use AD in reverse mode. To illustrate this point, we proceed to compute all second and cross derivatives i.e. \!the following Hessian matrix for a standard European Call option: 
		\begin{equation}
			\displaystyle 
			\begin{pmatrix}
				\displaystyle\frac{
				\partial ^2V}{
				\partial X_0^2} & \displaystyle\frac{
				\partial ^2 V}{
				\partial v 
				\partial X_0} & \displaystyle\frac{
				\partial ^2 V}{
				\partial r 
				\partial X_0} & \displaystyle\frac{
				\partial ^2 V}{
				\partial T 
				\partial X_0} \\
				
				\displaystyle\frac{
				\partial ^2V}{
				\partial X_0 
				\partial \sigma} &\displaystyle \frac{
				\partial ^2 V}{
				\partial \sigma ^2 } &\displaystyle \frac{
				\partial ^2 V}{
				\partial v 
				\partial r} &\displaystyle \frac{
				\partial ^2 V}{
				\partial T 
				\partial v} \\
				
				\displaystyle\frac{
				\partial ^2V}{
				\partial X_0 
				\partial r} &\displaystyle \frac{
				\partial ^2 V}{
				\partial v 
				\partial r} &\displaystyle \frac{
				\partial ^2 V}{
				\partial r^2} & \displaystyle\frac{
				\partial ^2 V}{
				\partial T 
				\partial r} \\
				
				\displaystyle\frac{
				\partial ^2V}{
				\partial X_0 
				\partial T} &\displaystyle \frac{
				\partial ^2 V}{
				\partial v 
				\partial T} &\displaystyle \frac{
				\partial ^2 V}{
				\partial r 
				\partial T} &\displaystyle \frac{
				\partial ^2 V }{
				\partial T^2 } \\
			\end{pmatrix}.
		\end{equation}
		
It is easily seen  that a Finite Difference procedure will require 36 (at least 33) evaluations of the original pricing function whereas we only call this function once if AD is used in reverse mode. Furthermore, we have to handle 4 different perturbation parameters.
		
The parameters are $X_0=90,\  K=100,\  \sigma =0.2, \ r=0.05$ and $T=1$ year.  The parameters of Monte Carlo are set to $200,000$ simulation paths and $50$ time steps. We used the library \texttt{adept 1.0} for the reverse mode. One great aspect here is that we only have one formula in the computer program to compute all the greeks, consequently one has just to specify which parameters are taken as variable for differentiation.

The results are shown  in the table \ref{cpuHessian}, clearly the reverse automatic differentiation combined with Vibrato is almost 4 times faster than the finite difference procedures. 
		\begin{table}
			[ht] 
			\begin{center}
				\begin{tabular}{|c|c|c|}\hline
					 \textbf{Mode} & \textbf{FD (MC)} & \textbf{VRAD (MC)} \\
					\hline \textbf{Time (sec)} & 2.01 & 0.47 
\\ \hline				\end{tabular}
				\caption{\label{cpuHessian}CPU time (in seconds) to compute the Hessian matrix of a standard European Call option (considering $X_0,\  \sigma,\  r,\  T$ as variables) in the Black-Scholes model.} 
			\end{center}
		\end{table}
	
		\section{Malliavin Calculus and Likelihood Ratio Method } \label{MALLIAVIN}
Here, we want to point out that Malliavin calculus and LRM are excellent methods but they have their own numerical issues especially with short maturities which may make VAD more attractive for a general purpose software.

 Let us start by recalling briefly the foundations of Malliavin calculus (further details are available in Nualart \cite{Nua06}, Fourni\'e et al.\cite{FLLL01} and in Gobet et al. \cite{GE05}, for instance). We recall the  Bismut-Elworthy-Li  formula (see  \cite{BEL98}, for example):
\begin{proposition}{\rm (Bismut-Elworthy-Li formula)}
Let $X$ be a diffusion process given by (\ref{1e}) with $d=1$, $b$ and $\sigma$  in $\mathcal{C}^1$ . Let $f:\mathbb{R} \rightarrow \mathbb{R}$ be $\mathcal{C}^1$  with 
$\mathbb{E} [ f(X_T)^2]$ and $\E[f'(X_T)^2]$ bounded.
Let $(H_t)_{t\in[0,T]}$ an  $\mathcal{F}$-progressively measurable process  in $L^2( [0,T] \times \Omega, dt \otimes d\mathbb{P})$ such that $ 
\mathbb{E}\left[\int _0^T H^2_sds \right]$ is finite.
Then
\begin{equation}
 \mathbb{E}\left[f(X_T)\int _ 0^T H_s dW_s \right] = \mathbb{E} \left[ f'(X_T)Y_T\int _0^T \frac{\sigma(X_s)H_s}{Y_s}ds\right]
\end{equation}
where $\displaystyle Y_t=\frac{dX_t}{dx}$ is the tangent process defined in (\ref{3e}). 
\end{proposition}
By choosing $H_t=Y_t/\sigma(X_t)$  the above yields
\begin{equation}
 \frac{\partial}{\partial x}\mathbb{E}\left[f(X^x_T)\right] = \mathbb{E} \left[ f(X^x_T )\underbrace{{\frac{1}{T} \int _0^T \frac{Y_s}{\sigma(X^x_s)} dW_s}}_\text{Malliavin weight}\right]
\end{equation}
provided  $f$  has polynomial growth and 
$\displaystyle \mathbb{E}\left[ \int _0^T \left( \frac{Y_t}{\sigma (X^x_t)} \right) ^2 \right] $ is finite.
%
\paragraph{Second Derivative.} In the context of the Black-Scholes model,  the Malliavin weights, $\pi_\Gamma$, for the Gamma is (see  \cite{Ben03}):
 \begin{equation}
 \pi_\Gamma =  \frac{1}{X_0^2\sigma T}\left( \frac{W^2_T}{\sigma T} - \frac{1}{\sigma} - W_T \right).
 \end{equation}
 Hence 
 \begin{equation}
 \Gamma_\text{Mal} = e^{-rT}\mathbb E \left[ (X_T-K)^+  \frac{1}{X_0^2\sigma T}\left( \frac{W^2_T}{\sigma T} - \frac{1}{\sigma} - W_T \right) \right].
 \end{equation}
The pure likelihood ratio method gives a similar formula (see Lemma \ref{LRM}) 
		\begin{equation}
			\Gamma_\text{\rm LR}=e^{-rT}\mathbb E\left[(X_T-K)^+\left( \frac{Z^2-1}{X_0^2\sigma ^2 T}- \frac{Z}{X_0^2\sigma \sqrt{T}}\right) \right].
		\end{equation}
LRPW is an improvement of LRM obtained by combining it with a pathwise method \cite{Gla03}. 
\begin{equation}
  \Gamma_\text{LRPW}=\frac{\partial}{\partial X_0}\left( e^{-rT}\mathbb E \left[ (X_T-K)^+\frac{Z}{X_0\sigma \sqrt{T}}\right] \right) =e^{-rT}\frac{K}{X_0^2\sigma \sqrt{T}}\mathbb{E}[Z\mathbf{1}_{\{X_T>K\}}].
\end{equation}
LRPW  is much cheaper than VAD, Malliavin or LRM and it is also less singular at $T=0$.  However all these methods require new analytically derivations for each new problem.
\subsection{Numerical Tests}
We compared VAD with LRPW and Malliavin calculus. The results are shown on Table \ref{table9}
\begin{table}
			[ht] 
			\begin{center}
				\begin{tabular}
					{|c|c|c|c|c|}\hline \textbf{T} & \textbf{VAD (MC)} & \textbf{FD (MC)} & \textbf{LRPW (MC)}& \textbf{Malliavin (MC)} \\
					\hline 1.00$e$+0 & 3.63$e$-5 & 1.76$e$-4&3.40$e$-4 & 9.19$e$-3 \\
						 5.00$e$-1  & 8.55$e$-5 & 3.11$e$-4&7.79$e$-4 & 1.62$e$-2\\
						 1.00$e$-1  & 6.64$e$-4 & 1.50$e$-3&4.00$e$-3 & 6.54$e$-2 \\
						 5.00$e$-2  & 1.49$e$-3 & 2.80$e$-3&7.51$e$-3 & 1.21$e$-1\\
						 1.00$e$-2  & 8.78$e$-3 & 1.84$e$-2&3.76$e$-2& 5.44$e$-1 \\
						 5.00$e$-3  &  1.86$e$-2 & 3.95$e$-2&7.55$e$-2 & 1.10$e$+0  \\
						 1.00$e$-3  &  9.62$e$-2& 1.77$e$-1&3.76$e$-1&5.74$e$+0 \\
						 5.00$e$-4  & 1.85$e$-1& 3.34$e$-1&7.56$e$-1& 1.07$e$+1 \\
						 1.00$e$-4  & 1.01$e$+0& 1.63$e$+0&3.77$e$+0& 5.26$e$+1 \\
						 5.00$e$-5 &  1.98$e$+0&3.46$e$+0&7.54$e$+0&1.09$e$+2 \\
						 1.00$e$-5 & 1.03$e$+1&1.78$e$+1&3.79$e$+1&5.40$e$+2
						 
\\ \hline				\end{tabular}
				\caption{\label{table9}{\it Variance of the Gamma of a standard European Call with short maturities in the Black-Scholes model. Gamma is computed with VAD, FD, LRPW and Malliavin. The computation are done on the same samples.}} 
			\end{center}
		\end{table}

\medskip

		The Gamma is computed with the same parameters as in the section \ref{NUMTEST}. The maturity is varying from $T=1$  to $10^{-5}$ year. The Monte Carlo parameters are also set to $100,000$ simulation paths and $25$ time steps.
		
Notice the inefficiency of LRPW, Malliavin Calculus and to a lesser degree of VAD and  Finite Difference when $T$ is small.

				\subsection*{Note on CPU} Tests have been done on an Intel(R) Core(TM) i5-3210M Processor @ 2,50 GHz. The processor has turbo speed of 3.1 GHz and two cores. We did not use parallelization  in the code.
\section{Conclusion} 
This article extends the work of Mike Giles and investigates the Vibrato method for higher order derivatives in quantitative finance. 

For a general purpose software Vibrato of Vibrato is too complex but we showed that it is essentially similar to the analytical differentiation of Vibrato. Thus AD of Vibrato is both general, simple and essentially similar to 	Vibrato of Vibrato of second derivatives.  	
We have also shown that Automatic differentiation can be enhanced to handle the singularities of the payoff functions of finance. While AD for second derivatives is certainly the easiest solution, it is not the safest and it requires an appropriate choice for the approximation of the Dirac mass.

Finally we compared with Malliavin calculus and LRPW.

 The framework proposed is easy to implement, efficient, faster and more stable than its competitors and does not require analytical derivations if local volatilities or payoffs are changed. 
 
  Further developments are in progress around nested Monte Carlo and  Multilevel-Multistep Richardson-Romberg extrapolation \cite{LP14} (hence an extension to \cite{BG11}).

		\subsection*{Acknowledgment} This work has been done with the support of ANRT and Global Market Solution inc. with special encouragements from Youssef Allaoui and Laurent Marcoux.

\bibliography{ref2}{} 
\bibliographystyle{plain}

		\begin{sidewaystable}
			[htbp] \caption{ \label{7dTab}Results for the price, the Delta and the Gamma of a Basket Option priced with the moment-matching approximation (reference values), Finite Difference on Monte Carlo and Vibrato plus Automatic Differentiation on Monte Carlo. The settings of Monte Carlo simulation are $1$ time step and $1,000,000$ simulation paths.}
			
			
			\bigskip \centering\small\setlength\tabcolsep{3pt}
			
			\hspace*{0cm}
			\begin{tabular}
				{c c c | c c c c c c c c c c c} \toprule ~$d$ ~&~ $T$ & ~ & 
				\begin{tabular}
					[x]{@{}c@{}}Price\\AMM
				\end{tabular}
				&
				\begin{tabular}
					[x]{@{}c@{}}Price\\(MC)
				\end{tabular}
				& ~~ & 
				\begin{tabular}
					[x]{@{}c@{}}Delta\\AMM
				\end{tabular}
				&
				\begin{tabular}
					[x]{@{}c@{}}Delta\\Vibrato (MC) 
				\end{tabular}
				& 
				\begin{tabular}
					[x]{@{}c@{}}Delta\\FD (MC)
				\end{tabular}
				& ~~ & 
				\begin{tabular}
					[x]{@{}c@{}}Gamma\\
					AMM
				\end{tabular}
				& 
				\begin{tabular}
					[x]{@{}c@{}}Gamma\\VAD (MC)
				\end{tabular}
				& 
				\begin{tabular}
					[x]{@{}c@{}}Gamma\\FD (MC)
				\end{tabular}
				\\
				
				\midrule 1 & 0.1& & 38.4285&37.3823 &~~& 0.55226& 0.55146&0.55423 & ~~&4.65557$e$-3& 4.66167$e$-3& 4.64998$e$-3\\
				2 & 0.1& &34.4401 &34.1232&~~& 0.27452 &0.27275&0.28467&~~ & 1.28903$e$-3& 1.34918$e$-3& 1.28193$e$-3\\
				3 & 0.1& & 46.0780&45.9829&~~ & 0.18319 &0.18220&0.18608& ~~& 4.29144$e$-4& 4.28572$e$-4& 4.21012$e$-4\\
				4 & 0.1& & 59.6741&58.7849 &~~& 0.13750&0.13639& 0.14147&~~ & 1.86107$e$-4& 1.93238$e$-4& 1.79094$e$-4\\
				5 & 0.1& & 92.8481&90.9001 &~~&0.10974&0.10889& 0.10956&~~ & 7.64516$e$-5& 7.79678$e$-5& 7.59901$e$-5\\
				6 &0.1 & & 139.235&141.766 &~~&0.09128&0.09017& 0.09048&~~ & 3.54213$e$-5& 3.71834$e$-5& 3.41114$e$-5\\
				7 &0.1 & & 155.492&153.392 &~~&0.07820&0.07744& 0.07766&~~ & 2.31624$e$-5& 2.09012$e$-5& 2.18123$e$-5\\
				& & & & & & ~~ & & &~~ & & & &\\
				1 &1 & & 155.389&154.797 &~~& 0.66111 &0.66039 &0.67277 & ~~&1.30807$e$-3 &1.30033$e$-3 & 1.32812$e$-3\\
				2 &1 & & 135.441 &133.101 &~~& 0.32583 &0.32186 &0.32547 &~~ & 3.80685$e$-4& 3.86998$e$-4&3.83823$e$-4 \\
				3 & 1& & 181.935&182.642&~~& 0.21775 &0.21497 &0.21619 & ~~& 1.26546$e$-4&1.34423$e$-4 &1.24927$e$-4 \\
				4 &1 & & 234.985& 232.018&~~& 0.16304 &0.16055 &0.01610 &~~ &5.49161$e$-5 &5.62931$e$-5 &5.50990$e$-5 \\
				5 &1 & & 364.651& 363.363&~~&0.13023 &0.12780 &0.12804&~~ &2.25892$e$-5&2.38273$e$-5 &2.19203$e$-5 \\
				6 & 1& & 543.629& 540.870&~~&0.10794 &0.10477 &0.10489 &~~ &1.04115$e$-5 &8.99834$e$-6 & 1.13878$e$-5\\
				7 & 1& & 603.818& 607.231&~~& 0.92420 &0.08995 &0.89945 &~~ &6.87063$e$-6 &7.70388$e$-6 &7.22849$e$-6 \\
				
				\bottomrule 
			\end{tabular}
			\hspace*{0cm} \vspace{1cm} \caption{\label{7dTime}Time computing (in seconds) for the Gamma with Finite Difference on Monte Carlo and with Vibrato plus Automatic Differentiation on Monte Carlo simulation, dimension of the problem are varying. The settings of Monte Carlo algorithm are the same as above.}
			
			
			\bigskip \centering\small\setlength\tabcolsep{3pt}
			
			\hspace*{0cm}
			\begin{tabular}
				{c| c c c c c c c c c c c c c c c} \toprule Method (Computing Gamma) &~~ & $d=1$ & ~~ &$d=2$ & ~~ &$d=3$ &~~ & $d=4$ &~~ & $d=5$ &~~ & $d=6$ &~~ & $d=7$ \\
				\midrule FD (MC) &~~ & 0.49 &~~& 0.95 &~~& 1.33 &~~ & 1.82 & ~~ & 2.26 & ~~ &2.91 & ~~ & 3.36\\
				VAD (MC) &~~ & 0.54 &~~ &0.77 &~~& 0.92 &~~& 1.21 &~~ & 1.50 & ~~ & 1.86 & ~~ &2.31 \\
				
				\bottomrule 
			\end{tabular}
			\hspace*{0cm} 
		\end{sidewaystable}		
		\newpage
		
				\begin{sidewaystable}
			[!h] \caption{\label{amctab} Results of the price, the Delta and the Gamma of an American option. The reference values are obtained via the Semi-Newton method plus Finite Difference, they are compared to Vibrato plus Automatic Differentiation on the Longstaff-Schwartz algorithm. We compute the standard error for each American Monte Carlo results. The settings of the American Monte Carlo are $50$ time steps and $50,000$ simulation paths.}
			
			
			\bigskip \centering\small\setlength\tabcolsep{3pt}
			
			\hspace*{0cm}
			\begin{tabular}
				{c c c | c c c c c c c c c c c } \toprule ~$S$ ~& ~~$\sigma$~ & ~$T$ ~~& 
				\begin{tabular}
					[x]{@{}c@{}}Price\\Ref. Value
				\end{tabular}
				& 
				\begin{tabular}
					[x]{@{}c@{}}Price\\(AMC)
				\end{tabular}
				& 
				\begin{tabular}
					[x]{@{}c@{}}Standard\\Error
				\end{tabular}
				& ~~ &
				\begin{tabular}
					[x]{@{}c@{}}Delta\\Ref. Value 
				\end{tabular}
				& 
				\begin{tabular}
					[x]{@{}c@{}}Delta\\Vibrato (AMC)
				\end{tabular}
				&
				\begin{tabular}
					[x]{@{}c@{}}Standard\\Error
				\end{tabular}
				& ~~ & 
				\begin{tabular}
					[x]{@{}c@{}}Gamma\\Ref. Value
				\end{tabular}
				& 
				\begin{tabular}
					[x]{@{}c@{}}Gamma\\VAD (AMC)
				\end{tabular}
				& 
				\begin{tabular}
					[x]{@{}c@{}}Standard\\Error
				\end{tabular}
				\\
				\midrule
				
				36 & 0.2 & 1 & 4.47919 & 4.46289 & 0.013 & ~~&0.68559&0.68123 & 1.820$e$-3&~~&0.08732 & 0.06745 & 6.947$e$-5\\
				36 & 0.2 & 2 &4.83852 & 4.81523 & 0.016& ~~&0.61860&0.59934 & 1.813$e$-3&~~&0.07381 & 0.06398& 6.846$e$-5 \\
				36 & 0.4 & 1 &7.07132 &7.07985 &0.016 &~~ &0.51019&0.51187 & 1.674$e$-3&~~&0.03305 & 0.03546& 4.852$e$-5\\
				36 & 0.4 & 2 &8.44139 &8.45612 &0.024 &~~ &0.44528&0.44102& 1.488$e$-3 &~~&0.02510 & 0.02591& 5.023$e$-5\\
				
				& & & & & & & & & & & & \\
				
				38 & 0.2 & 1 & 3.24164& 3.23324&0.013 &~~ & 0.53781&0.53063 & 1.821$e$-3&~~ &0.07349 & 0.07219 & 1.198$e$-4\\
				38 & 0.2 & 2 & 3.74004& 3.72705&0.015&~~ & 0.48612&0.46732 & 1.669$e$-3&~~ &0.05907 & 0.05789& 1.111$e$-4\\
				38 & 0.4 & 1 & 6.11553& 6.11209& 0.016&~~ & 0.44726&0.45079 & 1.453$e$-3&~~ &0.02989 & 0.03081& 5.465$e$-5\\
				38 & 0.4 & 2 & 7.59964& 7.61031& 0.025&~~ & 0.39786&0.39503 & 1.922$e$-3&~~ &0.02233 & 0.02342& 4.827$e$-5\\
				
				& & & & & & & & & & & & \\
				
				40 & 0.2 & 1 &2.31021& 2.30565&0.012&~~&0.41106& 0.40780& 1.880$e$-3& ~~& 0.06014& 0.05954& 1.213$e$-4 \\
				40 & 0.2 & 2 & 2.87877& 2.86072&0.014&~~&0.38017& 0.39266& 1.747$e$-3& ~~& 0.04717& 0.04567& 5.175$e$-4\\
				40 & 0.4 & 1 & 5.27933& 5.28741&0.015&~~&0.39051& 0.39485& 1.629$e$-3& ~~& 0.02689& 0.02798& 1.249$e$-5\\
				40 & 0.4 & 2 & 6.84733& 6.85873&0.026&~~&0.35568& 0.35446& 1.416$e$-3& ~~& 0.01987& 0.02050& 3.989$e$-5\\
				
				& & & & & & & & & & & & \\
				
				42 & 0.2 & 1 & 1.61364& 1.60788&0.011&~~&0.30614& 0.29712& 1.734$e$-3& ~~& 0.04764& 0.04563& 4.797$e$-5\\
				42 & 0.2 & 2 & 2.20694& 2.19079&0.014&~~&0.29575& 0.28175& 1.601$e$-3& ~~& 0.03749& 0.03601& 5.560$e$-5\\
				42 & 0.4 & 1 & 4.55055& 4.57191&0.015&~~&0.33973& 0.34385& 1.517$e$-3& ~~& 0.02391& 0.02426&3.194$e$-5\\
				42 & 0.4 & 2 & 6.17459& 6.18424&0.023&~~&0.31815& 0.29943& 1.347$e$-3& ~~& 0.01768& 0.01748& 2.961$e$-5\\
				
				& & & & & & & & & & & & \\
				
				44 & 0.2 & 1 & 1.10813&1.09648 &0.009&~~&0.21302& 0.20571& 1.503$e$-3& ~~& 0.03653& 0.03438& 1.486$e$-4\\
				44 & 0.2 & 2 & 1.68566&1.66903 &0.012&~~&0.22883& 0.21972& 1.487$e$-3& ~~& 0.02960& 0.02765& 2.363$e$-4\\
				44 & 0.4 & 1 & 3.91751&3.90838 &0.015&~~&0.29466& 0.29764& 1.403$e$-3& ~~& 0.02116& 0.02086& 1.274$e$-4\\
				44 & 0.4 & 2 & 5.57268&5.58252 &0.028&~~&0.28474& 0.28447& 1.325$e$-3& ~~& 0.01574& 0.01520& 2.162$e$-4\\
				
				\bottomrule 
			\end{tabular}
			\hspace*{0cm} 
		\end{sidewaystable}

		\end{document}